\documentclass[english,american]{elsart}
\usepackage[T1]{fontenc}
\usepackage{float}
\usepackage{amsmath}
\usepackage{graphicx}
\usepackage{amssymb}

\makeatletter

\usepackage{graphicx}

\DeclareMathOperator{\Pre}{Pre}

\newcommand{\eps}{\varepsilon}
\newcommand{\mT}{-}

\newcommand{\mM}{-}
\newcommand{\pM}{+}
\newcommand{\PP}{P}

\usepackage{babel}
\makeatother
\begin{document}
\begin{frontmatter}

\title{Designing Complex Networks}

\author{Raoul-Martin Memmesheimer$^{1,2,3}$ and Marc Timme$^{1,2,4}$}

\address{$^{1}$Network Dynamics Group, Max Planck Institute\\
for Dynamics \& Self-Organization  (MPIDS), }

\address{$^{2}$Bernstein Center for Computational Neuroscience (BCCN) Göttingen,}

\address{$^{3}$Fakultät für Physik, Georg-August-Universität Göttingen,\\
Bunsenstr.~10, 37073 Göttingen, Germany;}

\address{$^{4}$Center for Applied Mathematics, Theoretical and Applied Mechanics,
\\
Kimball Hall, Cornell University, Ithaca, NY 14853, USA.}

\ead{\textbf{memmesheimer@ds.mpg.de, marc.timme@ds.mpg.de}}

\begin{abstract}
We suggest a new perspective of research towards understanding the
relations between structure and dynamics of a complex network: Can
we design a network, e.g. by modifying the features of units or interactions,
such that it exhibits a desired dynamics? Here we present a case study
where we positively answer this question analytically for networks
of spiking neural oscillators. First, we present a method of finding
the set of all networks (defined by all mutual coupling strengths)
that exhibit an arbitrary given periodic pattern of spikes as an invariant
solution. In such a pattern all spike times of all the neurons are
exactly predefined. The method is very general as it covers networks
of different types of neurons, excitatory and inhibitory couplings,
interaction delays that may be heterogeneously distributed, and arbitrary
network connectivities. Second, we show how to design networks if
further restrictions are imposed, for instance by predefining the
detailed network connectivity. We illustrate the applicability of
the method by examples of Erd\"{o}s-R\'{e}nyi and power-law random
networks. Third, the method can be used to design networks that optimize
network properties. To illustrate the idea, we design networks that
exhibit a predefined pattern dynamics and at the same time minimize
the networks' wiring costs. 
\end{abstract}
\end{frontmatter}

\section{How does network structure relate to dynamics?}

Our understanding of complex systems, in particular biological ones,
ever more relies on mathematical insights resulting from modeling.
Modeling a complex system, however, is a highly non-trivial task,
given that many factors such as strong heterogeneities, interaction
delays, or hierarchical structure often occur simultaneously and thus
complicate mathematical analysis.

Many such systems consist of a large number of units that are at least
qualitatively similar. These units typically interact on a network
of complicated connectivity. Important example systems range from
gene regulatory networks in the cell and networks of neurons in the
brain to food webs of species being predator or prey to certain other
species \cite{AY92,HMIC01,DK02}.

A major question is how the connectivity structure of a network relates
to its dynamics and its functional properties. Researchers therefore
are currently trying to understand which kinds of dynamics result
from specific network connectivities such as lattices and random networks
as well as networks with small-world topology or power-law degree
distribution. \cite{S01,N03,S04}

Here we suggest a complementary approach: \emph{network design}. Can
we modify structural features of a complex network such that it exhibits
a desired dynamics? We positively answer this question analytically
for a class of spiking neural network models and illustrate our findings
by numerical examples.

In neurophysiological experiments, recurring patterns of temporally
precise and spatially distributed spiking dynamics have been observed
in different neuronal systems \emph{in vivo} and \emph{in vitro} \cite{A93,S99,I04,GS05}.
These spike patterns correlate with external stimuli (events) and
are thus considered key features of neural information processing
\cite{A04}. Their dynamical origin, however, is unknown. One possible
explanation for their occurrence is the existence of excitatorily
coupled feed-forward structures, synfire chains \cite{A82,HHP95,DGA99,AMAH03},
\textbf{}which are embedded in a network of otherwise random connectivity
and receive a large number of random external inputs. Such stochastic
models explain the recurrence of coordinated spikes but do not account
for the specific relative spike times of individual neurons, although
these are discussed to be essential for computation, too. To reveal
mechanisms underlying specific spike patterns and their computational
capabilities, our long term aim is to develop and analyze a new, deterministic
network model that explains the occurrence of specific precisely timed
spike patterns exhibiting realistic features. The work presented here
constitutes one of the first steps in this direction \textbf{}(cf.
also \cite{J02,BR02,DTDWG04}) and focuses on designing networks such
that they exhibit an arbitrary predefined periodic spike pattern. 

The article is organized as follows. In section 2 we introduce a class
of network models of spiking neurons and illustrate their relation
to standard modeling approaches using differential equations. In section
3 we design networks by deriving systems of equations and inequalities
that analytically restrict the set of networks (in the space of all
coupling strengths) such that they exhibit an arbitrary predefined
periodic spike pattern as an invariant dynamics. It turns out that
such systems are often underdetermined such that further requirements
on the individual units, the interactions and the network connectivity
can be imposed. We illustrate this in section 4 by specifying completely,
for each neuron, the sets of other neurons it receives spikes from,
i.e. the entire network connectivity. We present examples of networks
with specified connectivities of different statistics and design their
coupling strengths such that they exhibit the same spike pattern.
In section 5 we demonstrate the possibility of designing networks
that are optimal (with respect to some cost function). We present
illustrating examples of networks that exhibit a certain pattern of
precisely timed spikes and at the same time minimize wiring costs.
In section 6 we provide a brief step-by-step instruction for applying the presented method. 
Section 7 provides the conclusions and highlights open questions regarding
the design of complex networks.

The method of finding the set of networks exhibiting a predefined
pattern (parts of sections 2 and 3 of this article) was briefly reported
before in reference \cite{MT05} and in abstract form in \cite{MT06CNS},
where only the case of non-degenerate patterns, identical delays and
identical neurons was treated explicitely. Small inhomogeneities have
been discussed in \cite{DTDWG04}. Here we include also degenerate
patterns, heterogeneously distributed delays and allow for different
neuron types. Moreover, we present new applications of network design,
see in particular sections 4 and 5.

\section{Model neural networks}

\subsection{Phase model\label{sub:Phase-model}}

\begin{figure}
\begin{center}\includegraphics[%
  width=100mm]{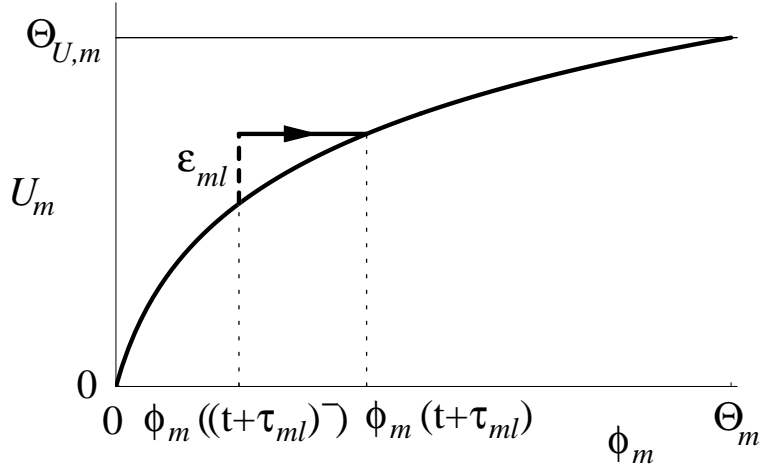}\end{center}

\caption{Phase dynamics in response to incoming excitatory spike. The rise
function $U_{m}$ of neuron $m$ is plotted as a function of its phase
$\phi_{m}$. In the absence of interactions, $\phi_{m}(t)$ increases
uniformly with time $t$ according to Eq.\ (\ref{eq:PhiDotOne}). If
a spike is sent by neuron $l$ at time $t$, it is received by neuron
$m$ at time $t+\tau_{ml}$ and induces a phase jump $\phi_{m}((t+\tau_{ml})^{-})\rightarrow\phi_{m}(t+\tau_{ml})$
that is mediated by the rise function $U_{m}$ and its inverse according
to (\ref{eq:PhaseInteraction}) and (\ref{eq:HintermsofU}). Here
$\Theta_{U,m}=U_{m}(\Theta_{m})$ is the threshold for the membrane
potential, cf. sec. \ref{sub:Phases-vs.-neural}. \label{cap:PhaseDynamics}}
\end{figure}

Consider a network of $N$ oscillatory neurons that interact by sending
and receiving spikes via directed connections. The network connectivity
is arbitrary and defined if we specify for each neuron $l\in\{1,\ldots,N\}$
the sets $\Pre(l)$ from which it receives input connections. One
phase-like variable $\phi_{l}(t)$ specifies the state of each neuron
$l$ at time $t$. A continuous strictly monotonic increasing rise
function $U_{l}$, $U_{l}(0)=0$, defines the membrane potential $U_{l}(\phi_{l})$
of the neuron, representing its subthreshold dynamics \cite{MS90},
see Fig. \ref{cap:PhaseDynamics}. The neurons interact at discrete
event times when they send or receive spikes. We first introduce the
model for non-degenerate events, i.e. non-simultaneous event times,
and provide additional conventions for degenerate events in the next
subsection.
 \textbf{}

In the absence of interactions, the phases increase uniformly obeying
\begin{equation}
d\phi_{l}/dt=1.\label{eq:PhiDotOne}\end{equation}
 When $\phi_{l}$ reaches the (phase-)threshold of neuron $l$, $\phi_{l}(t^{-})=\Theta_{l}>0$,
it is reset, $\phi_{l}(t)=0$, and a spike is emitted. After a delay
time $\tau_{ml}$ this spike signal reaches the post-synaptic neuron
$m$, inducing an instantaneous phase jump\begin{equation}
\phi_{m}\left(t+\tau_{ml}\right)=H_{\varepsilon_{ml}}^{(m)}\left(\phi_{m}\left(\left(t+\tau_{ml}\right)^{-}\right)\right),\label{eq:PhaseInteraction}\end{equation}
 mediated by the continuous response function \begin{equation}
H_{\varepsilon}^{(m)}(\phi)=U_{m}^{-1}(U_{m}(\phi)+\varepsilon)\label{eq:HintermsofU}\end{equation}
 that is strictly monotonic increasing, both as a function of $\varepsilon$
and of $\phi$. Here, $\varepsilon_{ml}$ denotes the strength of
the coupling from neuron $l$ to $m$. This coupling is called inhibitory
if $\eps_{ml}<0$ and excitatory if $\eps_{ml}>0$. We note that sending
and receiving of spikes are the only nonlinear events occurring in
these systems. Throughout the manuscript,  $\phi_{l}(t)$ is assumed to be 
piecewise linear for all $l$ such that in any finite time interval there are
only a finite number of spike times.

\subsection{Degenerate event timing \label{subsection:Degenerate-event-timing}}

These events of sending and receiving spikes might sometimes occur
simultaneously such that care has to be taken in the definition of
the model dynamics. Simultaneous events occurring at different neurons
do not cause any difficulties because an arbitrary order of processing
does not affect the collective dynamics at any future time. However,
if two or more events occur simultaneously at the same neuron, we
need to specify a convention for the order of processing. We will
therefore go through the possible combinations in the following: 

(i) \emph{spike sending due to spike reception}: The action of a received
spike might be strong enough such that the excitation is supra-threshold,\begin{equation}
U_{m}\left(\phi_{m}\left(\left(t+\tau_{ml}\right)^{-}\right)\right)+\varepsilon_{ml}\geq U_{m}\left(\Theta_{m}\right).\label{eq:SupraThrInput}\end{equation}
We use the convention that neuron $m$ sends a spike simultaneous
to the reception of another spike from neuron $l$ at time $t+\tau_{ml}$
and is reset to

\begin{equation}
\phi_{m}\left(t+\tau_{ml}\right)=0.\label{eq:SupraThrInputReset}\end{equation}

(ii) \emph{spike received at sending time}: If neuron $m$ receives
a spike from neuron $l$ exactly at the same time when $m$ was about
to send a spike anyway, \begin{equation}
\phi_{m}\left(\left(t+\tau_{ml}\right)^{-}\right)=\Theta_{m},\label{eq:PhaseFreeToThreshold}\end{equation}
 we take the following convention for the order processing: first
the spike is sent and the phase is reset to zero, then the spike is
received such that \begin{equation}
\phi_{m}\left(t+\tau_{ml}\right)=H_{\varepsilon_{ml}}^{(m)}\left(0\right).\label{eq:NonzeroAfterReset}\end{equation}
 If the spike received causes again a supra-threshold excitation,
we neglect a second spike potentially generated at time $t+\tau_{ml}$ and
just reset the neuron $m$ to zero as in (\ref{eq:SupraThrInputReset}). 

(iii) \emph{simultaneous reception of multiple spikes}: If multiple
spikes are received simultaneously by the same neuron and each subset
of spikes does \emph{not} cause a supra-threshold excitation (as in
(\ref{eq:SupraThrInput})), a convention about the order of treatment
is not necessary as can be seen from the following argument. If neuron
$m$ at time $\theta$ simultaneously receives $h\in\mathbb{{N}}$
spikes from neurons $l_{1},...,l_{h\,}$, and $\sigma:\{1,...,h\}\rightarrow\{1,...,h\}$
is an arbitrary permutation of the first $h$ integers, we have

\begin{align}
 & H_{\eps_{ml_{\sigma(1)}}}^{(m)}(H_{\eps_{ml_{\sigma(2)}}}^{(m)}(...H_{\eps_{ml_{\sigma(h)}}}^{(m)}(\phi_{m}(\theta^{-}))...))\nonumber \\
= & U_{m}^{-1}[U_{m}(U_{m}^{-1}[U_{m}(...U_{m}^{-1}[U_{m}(\phi_{m}(\theta^{-}))+\eps_{ml_{\sigma(h)}}]...)+\eps_{ml_{\sigma(2)}}])+\eps_{ml_{\sigma(1)}}]\nonumber \\
= & U_{m}^{-1}[U_{m}(\phi_{m}(\theta^{-}))+\eps_{ml_{\sigma(h)}}+...+\eps_{ml_{\sigma(2)}}+\eps_{ml_{\sigma(1)}}]\nonumber \\
= & H_{\eps_{ml_{1}}+\eps_{ml_{2}}+...+\eps_{ml_{h}}}^{(m)}(\phi_{m}(\theta^{-})).\label{eq:Hcomposition}\end{align}

Treating the incoming spikes separately in arbitrary order is therefore
equivalent to treating them as one spike from a hypothetic neuron
with coupling strength $\eps_{ml_{1}}+\eps_{ml_{2}}+...+\eps_{ml_{h}}$
to neuron $m$. Moreover, upon sufficiently small changes of the spike
reception times, the sub-threshold response of a neuron $m$ continuously
changes with these reception times, even if their order changes: For every ordering of the reception times, the total phase response converges, in the limit of identical times, to the phase response to simultaneously received spikes. 
This is because the neuron's response function $H^{(m)}$ is identical for different incoming spikes. We note that this might not be the case in neurobiologically
more realistic models if they take into account that spikes from different
neurons arrive at differently located synapses. These spikes may have
a different effect on the postsynaptic neuron even if they generate
the same amount of charge flowing into (or out of) the neuron.

We extend the definition \begin{equation}
\phi_{m}(\theta)=H_{\eps_{ml_{1}}+\eps_{ml_{2}}+...+\eps_{ml_{h}}}^{(m)}(\phi_{m}(\theta^{-}))\label{eq:WellDefined}\end{equation}
 for the processing of multiple spike receptions to more involved
cases, where a subset of spikes generates a spike. Treating this subset
first would result in a different dynamics than summing up all couplings
strength, e.g. if the remaining couplings balance the strong excitatory
subset. In this case the order of treatment is not arbitrary and the
phase as well as the spikes generated in response to the receptions
do not continuously depend on the spike reception times; as a convention,
we sum the coupling strengths first, as in (\ref{eq:WellDefined}).

The generalization of (i) and (ii) to the case of multiple spikes
received simultaneously is straightforward. The dynamics however will
in general also \emph{not} depend continuously on the reception times. 

(iv) \emph{simultaneous sending of multiple spikes:} As we exclude
the simultaneous sending of multiple spikes by the same neuron, if
several spikes are sent simultaneously, they are sent by different
neurons; therefore no difficulties arise and we need no extra convention.

\subsection{Phases vs. neural membrane potentials\label{sub:Phases-vs.-neural}}

\begin{figure}
\begin{center}\includegraphics[width=140mm]{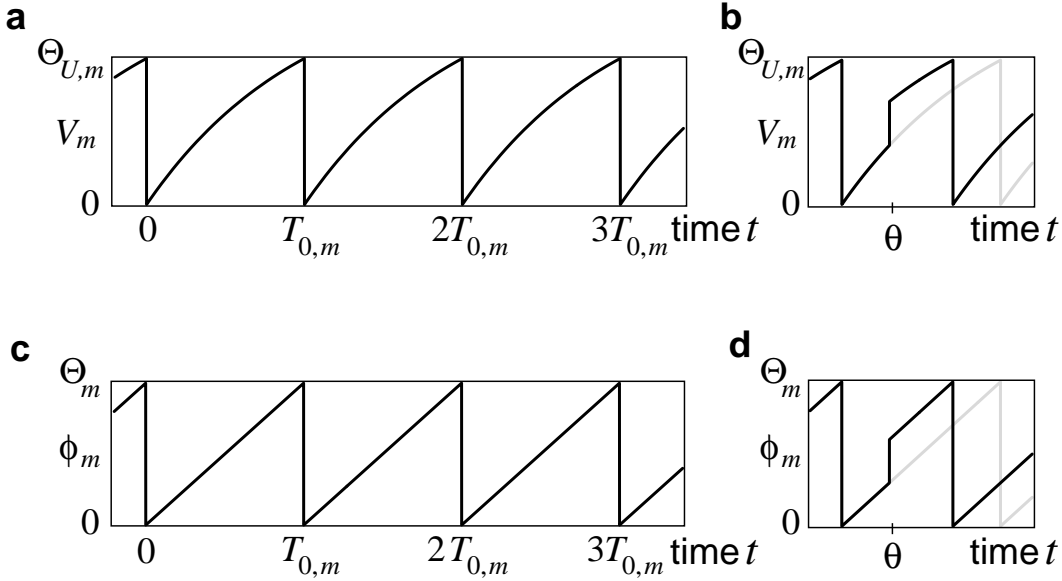}\end{center}

\caption{Relation between phase and membrane potential dynamics. (a,b) Dynamics
of membrane potential $V_{m}(t)$ of neuron $m$. (a) The free dynamics
is periodic with period $T_{0,m}$; (b) dynamics in response to an
incoming excitatory spike at time $\theta$. (c,d) Dynamics of
$\phi_{m}(t)$ representing a phase-like variable of the membrane potential
dynamics displayed in panels (a) and (b). (c) Periodic phase dynamics has
the same period $T_{0,m}$; (d) dynamics in response to input implies
phase jump given by Eq.~(\ref{eq:PhaseInteraction}). \label{cap:PotentialVphasephi}}
\end{figure}

The above phase dynamics in particular represent (cf.~also \cite{MS90,EPG95,T02,TWG03,TWG04})
dynamics of neural membrane potentials defined by a hybrid
dynamical system \cite{AT05} consisting of maps that occur at discrete
event times and ordinary differential equations, or, formally, of a differential
equation of the form
\begin{equation}
\frac{dV_{m}}{dt}=f_{m}(V_{m})+I_{m}(t).
\label{eq:V_dynamics}\end{equation}
 Here $I_{m}(t)=\sum_{l,n}\eps_{ml}\delta(t-t{}_{l,n}-\tau_{ml})$
is a sum of delayed $\delta$-currents induced by the neurons $l\in\Pre(m)$
sending their $n$th spike at time $t{}_{l,n}$. A solution $V_{m}(t)$
gives the membrane potential of neuron $m$ at time $t$ in response
to the current from the network $I_{m}(t)$. \textbf{}See Fig.~\ref{cap:PotentialVphasephi}
for an illustration. A spike is sent by neuron $m$ whenever a potential
threshold is crossed (for supra-threshold input, e.g., $V_{m}(t{}_{m,n}^{-})+\eps_{ml}\geq\Theta_{U,m}$
for some $l$; otherwise $V_{m}(t{}_{m,n}^{-})=\Theta_{U,m}$), leading
to an instantaneous reset of that neuron, $V_{m}(t{}_{m,n})=0$ (or
to a nonzero value equal to the coupling strength of the incoming
pulse, if a subthreshold spike reception coincides with the potential satisfying
$V_{m}(t{}_{m,n}^{-})=\Theta_{U,m}$, according to (ii) in sub-section
\ref{subsection:Degenerate-event-timing}). The positive function
$f_{m}(V)>0$ (for all admissible $V$) yields a solution $\tilde{V}_{m}(t)$
of the free ($I_{m}=0)$ dynamics that satisfies the initial
condition $\tilde{V}_{m}(0)=0$. We continue this solution $\tilde{V}_{m}$
on the real interval $t\in(B,\Theta_{m}]$, i.e. to negative real
arguments $t$ with infimum $B\in\mathbb{R}^{-}\cup\{-\infty\}$ and
to positive real $t$ until $\Theta_{m}\in\mathbb{R}^{+}$ where
$\tilde{V}_{m}(\Theta_{m})=\Theta_{U,m\,}$.
We note that a too large inhibition can be inconsistent with a possible lower
bound $\lim_{\phi\searrow B}\tilde{{V}}_{m}(\phi)>-\infty$ of the membrane
potential as present, e.g., for the leaky-integrate-and-fire neuron with
$\gamma<0$ (cf. Eq.\ (\ref{eq:UIF})). However, it does not change the methods
developed below using the
phase representation and is therefore not considered in the following.
The above rise function $U_{m}$ is then defined via $\tilde V_m$ as
\begin{equation}
U_{m}(\phi):=\tilde{V}_{m}(\phi),\label{eq:UV}\end{equation}
 where $\phi\in(B,\Theta_{m}]$. The potential dynamics can now
be expressed in terms of a natural phase $\phi_{m}(t)$ such that
\begin{equation}
V_{m}(t)=U_{m}(\phi_{m}(t))\label{eq:VandPhi}\end{equation}
 for all $t$. Since $\tilde{V}_{m}(t)$ is strictly monotonically
increasing in $t$, this also holds for $U_{m}(\phi)$ in $\phi$,
and the inverse $U_{m}^{-1}$ exists on the interval $(\lim_{\phi\searrow B}\tilde{{V}}_{m}(\phi),\Theta_{U,m}]$.
Therefore, the phase at the initial time, say $t_{0}$, can be computed
from the initial membrane potential via $\phi_{m}(t_{0})=U_{m}^{-1}(V_{m}(t_{0}))$.
If the dynamics evolves freely, the phase satisfies $d\phi_{m}/dt=1$,
and is reset to zero when its threshold $\Theta_{m}$ is reached,
cf. Fig.~\ref{cap:PotentialVphasephi}. Due to the invertibility
of $U_{m}$, there is a one-to-one mapping 

\begin{equation}
\Theta_{m}=U_{m}^{-1}\left(\Theta_{U,m}\right)\label{eq:ThresholdPhiU}\end{equation}
between the threshold $\Theta_{U,m}$ in the membrane potential and
the threshold $\Theta_{m}$ in the phase. This phase threshold equals
the free period of neuron $m$, \begin{equation}
\Theta_{m}=T_{0,m\,},\label{eq:ThetamEqualsTm}\end{equation}
 due to the constant unit velocity (\ref{eq:PhiDotOne}) of the phase
in the absence of input: starting from zero after reset, the phase
$\phi_{m}$ needs a time $\Theta_{m}$ to reach the threshold. Thus
$\Theta_{m}$ is the intrinsic inter-spike-interval and $1/\Theta_{m}$
is the intrinsic frequency of neuron $m$. 

In the presence of interactions, the size of the discontinuities in
the phase resulting from spike receptions have to match the size of
the corresponding discontinuities in the membrane potential, cf. Figs.
\ref{cap:PhaseDynamics} and \ref{cap:PotentialVphasephi}. To compute
the correct size, we first compute the membrane potential $V_{m}(\theta^{-})=U_{m}(\phi_{m}(\theta^{-}))$
of neuron $m$ just before the reception time $\theta$ of a spike
from neuron $l$. The membrane potential after the interaction is
given by $V_{m}(\theta)=U_{m}(\phi_{m}(\theta^{-}))+\varepsilon_{ml}$
due to (\ref{eq:V_dynamics}). We return to the phase representation
using the inverse rise function and compute the phase after the interaction
\begin{equation}
\phi_{m}(\theta)=U_{m}^{-1}(V_{m}(\theta))=U_{m}^{-1}(U_{m}(\phi_{m}(\theta^{-}))+\varepsilon_{ml})=H_{\varepsilon_{ml}}^{(m)}(\phi_{m}(\theta^{-})),\label{eq:VUUinvInteraction}\end{equation}
and arrive at relation (\ref{eq:PhaseInteraction}) between the phase
before and after interaction. Together with the fact that the reset
levels, the thresholds and the free dynamics match due to $U_{m}^{-1}(0)=0$,
Eqns.~(\ref{eq:ThresholdPhiU}) and (\ref{eq:UV}), this shows the
equivalence of the membrane potential dynamics given by the hybrid
system (\ref{eq:V_dynamics}) and the phase dynamics defined in section~\ref{sub:Phase-model}.

As an important example, the \emph{leaky integrate-and-fire neuron},
defined by $f_{m}(V)=I-\gamma V$, results in the specific form \begin{equation}
U_{\textrm{IF}}(\phi)=(I/\gamma)(1-e^{-\gamma\phi}).\label{eq:UIF}\end{equation}
Here $I>0$ is a constant external input and $\gamma\in\mathbb{{R}}$
specifies the dissipation in the system. For normal dissipation, $\gamma>0$,
$U_{\textrm{IF}}(\phi)$ is concave, $U_{\textrm{IF}}''(\phi)<0$,
bounded above by $I/\gamma$ and it approaches this value for $\phi\rightarrow\infty$.
Assuming $I/\gamma>\Theta_{U}$ we obtain an intrinsically oscillatory
neuron. For $\gamma<0,$ $U_{\textrm{IF}}(\phi)$ is convex, $U_{\textrm{IF}}''(\phi)>0$
, and bounded below by \textbf{$I/\gamma<0$}. It grows exponentially
with $\phi$ such that, apart from $\Theta_{U}>0$, no condition is
necessary to obtain a self-oscillatory neuron. For $\gamma=0$, the
dynamics of an isolated neuron is trivial and specified by $U_{\textrm{IF}}(\phi)=I\phi$.
The phase-threshold (\ref{eq:ThresholdPhiU}) for a particular integrate-and-fire
neuron $m$ is given by \begin{equation}
\Theta_{m}=U_{m}^{-1}(\Theta_{U,m})=\gamma_{m}^{-1}\ln(I_{m}/(I_{m}-\gamma_{m}\Theta_{U,m}))\label{eq:ThresholdPhiUIF}\end{equation}
 if the parameters are $I_{m}$ and $\gamma_{m}$; for $\gamma_{m}=0$
we have $\Theta_{m}=\Theta_{U,m}/I_{m}$, the limit $\gamma_{m}\rightarrow0$
in (\ref{eq:ThresholdPhiUIF}). 

Another interesting and analytically useful example is given by the biological
oscillator model first introduced by \emph{Mirollo and Strogatz} \cite{MS90},
\begin{equation}
U_{\textrm{MS}}(\phi)=b^{-1}\ln(1+a^{-1}\phi),\label{eq:UMS}\end{equation}
$ab>0$, which result from a differential equation (\ref{eq:V_dynamics})
with \\
$f_{m}(V)=\exp(-bV)/(ab)$. Here $U_{\textrm{MS}}(\phi)$ is concave
for $a,b>0$ and convex for $a,b<0$. In the former case the domain of 
$U_{\text{MS}}$ is $\phi\in(-a,\infty)$, with $U_{\textrm{MS}}(\phi)\rightarrow\infty$
as $\phi\rightarrow\infty$; in the latter case the domain is $\phi\in(-\infty,|a|)$,
where $U_{\text{MS}}(\phi)\rightarrow\infty$ as $\phi\nearrow|a|$.
Therefore, in both cases, there are no additional conditions on $\Theta_{U}$.
The threshold for the phase of a particular neuron $m$ is given by\begin{equation}
\Theta_{m}=U_{m}^{-1}(\Theta_{U,m})=a_{m}(\exp(b_{m}\Theta_{U,m})-1)\label{eq:ThresholdPhiUMS}\end{equation}
for parameters $a_{m},b_{m}$.

We note a direct relation between neural oscillators of leaky integrate-and-fire
and Mirollo-Strogatz type: the rise function of a Mirollo-Strogatz
oscillator is the inverse of the rise function of a leaky integrate-and-fire
neuron. For $x$ in the domain of $U_{\textrm{MS}}$ (or $U_{\textrm{IF}}^{-1}$) we have 
\begin{equation}
U_{\textrm{MS}}(x)=\frac{1}{b}\ln(1+\frac{x}{a})=-\frac{1}{\gamma}\ln(1-\frac{\gamma}{I}x)=U_{\textrm{IF}}^{-1}(x)\label{eq:UMSequalsUIFinv}\end{equation}
 when setting $b=-\gamma$, $a=-I/\gamma$. This can be directly verified
by explicitely inverting $U_{\textrm{IF}}$. To our knowledge, this
has not been noticed before but might be useful to establish
equivalences for dynamical properties of networks of such neurons
because the response function $H$ contains both, the rise function
$U$ and its inverse $U^{-1}$, cf. Eq.~(\ref{eq:HintermsofU}).

\begin{figure}
\begin{center}\includegraphics[%
  width=1.0\textwidth]{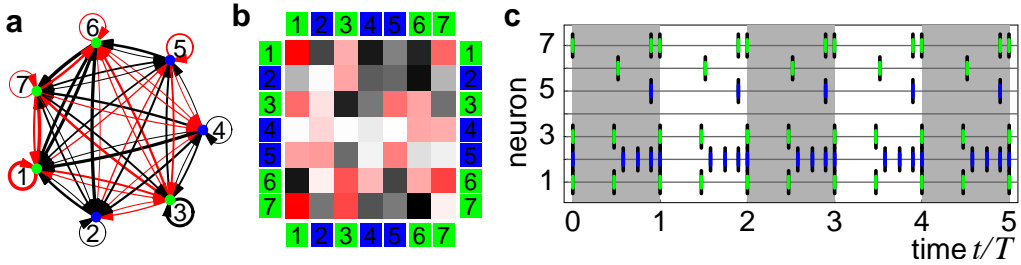}\end{center}

\caption{(color) Spike pattern in a small network ($N=7$). (a,b)
Network of four leaky integrate-and-fire (green) and three Mirollo-Strogatz
(blue) neurons in graph and matrix representation. The parameters
of the leaky integrate-and-fire neurons are randomly chosen within
$\gamma_{m}\in(0.5,1.5)$, $I_{m}=(1.08,2.08)$ and $\Theta_{m}\in(0.5,1.5)$.
(If $\gamma_{m}=1$ and $I_{m}=e/(e-1)\approx1.58$ as well as $\Theta_{m}=1$
then $\Theta_{U,m}=1$.) The parameters $b_{m}$ of the Mirollo-Strogatz
neurons are randomly chosen within $b_{m}\in(0.7,1.5)$, then $a_{m}$
is chosen within $a_{m}\in(1/(e^{b_{m}}-1)-0.1,1/(e^{b_{m}}-1)+0.1)$
and $\Theta_{m}\in(0.5,1.5)$. The delays are randomly distributed
within $\tau_{ml}\in(0.1,0.9)$. Connections are either excitatory
(black) or inhibitory (red) . In (a) the line widths of the links,
in (b) the color intensities are proportional to the coupling strengths.
The network is a realization randomly drawn from those networks with
couplings in the range $\eps_{lm}\in(-1.5,1.5)$ that exhibit the
predefined pattern displayed in (c) (black bars underlying the colored
ones). (c) The spiking dynamics (green and blue bars according to
neuron type) of the network shown in (a) and (b) perfectly agrees
with the predefined pattern of period $T=1.3$ (black bars). The pattern
includes several simultaneous spikes. One neuron, $l=4$, is silenced
(non-spiking). \label{cap:heterogeneity}}
\end{figure}

\section{Network Design: \protect \\
Analytically restricting the set of admissible networks\label{sec:NetworkDesign}}

In this section, we explain the underlying ideas of how to design
a network. For the class of systems introduced above, we derive conditions
on a network under which it exhibits an arbitrary predefined periodic
spike pattern. To avoid extensively many case distinctions, the following 
presentation requires that between any two subsequent spike times $t$  
and $t'$ of a  neuron $l$ that neuron receives at least one spike in the 
interval $(t,t')\cap (t,t+\Theta_l)$. This simply ensures that all spike 
times in a pattern can be modified by the coupling strengths.

\begin{defn}
\textbf{\emph{(Admissible Network)}} Given a predefined spike pattern,
we call a network that exhibits this pattern as an invariant dynamics
an \emph{admissible} network. 
\end{defn}
We assume here that all neuron parameters ($U_{m}$, $\Theta_{m}$)
and delay times $\tau_{ml}$ are given and fixed in a network; the
task is to find networks with these given features that exhibit a
desired spike pattern as an invariant dynamics. To design these networks,
we choose to vary the coupling strengths $\eps_{ml}$. It turns out
that there is often a family of solutions such that networks with
very different configurations of the coupling strengths are admissible;
below we derive analytical restrictions that define the set of all
networks exhibiting such a pattern. Of course there might be situations,
where other parameters, such as the delays \textbf{}\cite{EPECM99}
are desired to be variable as well (or only). The key aspects of the
approach presented below can be readily adapted to such design tasks. 

The analysis presented here is very general. It covers arbitrarily
large networks, different types of neurons, heterogeneously distributed
delays and thresholds (and thus intrinsic neuron frequencies), combinations
of inhibitory and sub- and supra-threshold excitatory interactions
as well as complicated pattern dynamics that include degenerate event
times, multiple spiking of the same neuron within the pattern and
silent neurons that never emit a spike. Figure \ref{cap:heterogeneity}
illustrates such a general case.

\subsection{Pattern Periodicity imposes restrictions}

Here we provide an indexing method for any given periodic spike pattern. We
then explain the relations between the periodicity of a spike pattern and the
possible) periodicity of a trajectory in state space along which an
appropriate network dynamical system generates that pattern.

What characterizes a periodic pattern of precisely timed spikes? Let
$t_{i'}$, $i'\in\mathbb{Z}$, be an ordered list of times at which
a neuron emits the $i'\textrm{th}$ spike occurring in the network,
such that $t_{j'}\geq t_{i'}$ if $j'>i'$. Assume a periodic pattern
consists of $M$ spikes. Such a pattern is then characterized by its
period $T$, by the times $t_{i}\in[0,T)$ of spikes $i\in\{1,...,M\}$
within the first period, and by the indices $s_{i}\in\{1,\ldots,N\}$
identifying the neuron that sends spike $i$ at $t_{i}\,$.
If two
or more neurons in the network simultaneously emit a spike, i.e. $t_{i}=t_{j}$
with $i\neq j$, the above order is not unique and we
fix the corresponding indices $s_{i}$
and $s_{j}$ arbitrarily.
\textbf{} The periodicity then entails
\begin{equation}
t_{i}+nT=t_{i+nM}\,\,\textrm{and}\,\, s_{i}=s_{i+nM},\label{eq:TemporalPeriodicity}\end{equation}
where $n\in\mathbb{Z}$ and the definition of $s$ was appropriately extended. This imposes conditions on the time evolution
of the neurons' phases. Suppose a specific neuron $l$ fires at $K(l)$
different times $t_{i_{k}}\in[0,T)$, $k\in\{1,...,K(l)\}$ within
the first period. \textbf{}For non-degenerate event times this implies
\begin{equation}
\phi_{l}(t_{i_{k}}^{-})=\Theta_{l},\label{eq:F1}\end{equation}
 for the neuron's spike times, whereas at any other time $t\in[0,T)$,
$t\neq t_{i_{k}}$ for all $k$, \begin{equation}
\phi_{l}(t^{-})<\Theta_{l},\label{eq:S1}\end{equation}
to prevent untimely firing.

Due to the periodicity of the pattern, we can assume
without loss of generality that the delay times $\tau_{ml}$
are smaller than the patterns period $T$; otherwise, we take them modulo $T$ without changing
the invariant dynamics such that $\tau_{ml}\in[0,T)$.

\begin{thm}\label{pattperiod}
The periodicity of the phases of all neurons in the network is sufficient
for the periodicity of the spiking times of each neuron. If
there are no supra-threshold excitations in the network, the spike pattern
has the period of the phase
dynamics.
\end{thm}

If the phase dynamics 
is periodic
with period $T$ and no supra-threshold excitations occur, it satisfies
in particular $\phi_{l}((t_{i_{k}}+nT)^{-})=\Theta_{l}$ and
$\phi_{l}((t+nT)^{-})<\Theta_{l}$ for $t_i \neq t_{i_k}$;
$t_{i_{k}}\in[0,T)$,
$k\in\{1,\ldots,K(l)\}$, are the firing times of neuron $l$ in the
first period. Therefore the sub-pattern of spikes generated by neuron
$l$ is periodic with period $T$. Since $l$ is arbitrary, the entire
pattern is periodic with period $T$. 

Interestingly, if there are supra-threshold excitations, the 
sub-pattern of a neuron need not have the period $T$ of the
phases, as can be seen from a simple, albeit constructed example:
Consider a neuron $l$, which is coupled only to itself and receives
input from itself as well as once per phase period $T$ from only
one other neuron $m$. If neuron $l$ receives a supra-threshold input
from neuron $m$ at time $\theta$, we have $\phi_{l}(\theta^{-})<\Theta_{l}$
and $U_{l}(\phi_{l}(\theta^{-}))+\eps_{lm}\geq U_{l}(\Theta_{l})$.
Suppose the delay of the coupling from $l$ to $l$ is $\tau_{ll}=T$,
i.e. equal to the period of the phases, and the coupling strength
$\eps_{ll}$ is inhibitory and such that $H_{\eps_{lm}+\eps_{ll}}^{(l)}(\phi_{l}(\theta^{-}))=0$,
i.e. $\eps_{ll}=-U_{l}(\phi_{l}(\theta^{-}))-\eps_{lm}<0$. Then the
phase of neuron $l$ can be periodic, whether or not it receives a
spike from itself because $\phi_{l}(\theta)=0$ in each case, either
due to the reset of neuron $l$ or due to the inhibitory spike received
from itself. Now, if neuron $l$ sent a spike at time $\theta$, there
will be no spike sending at $\theta+T$ because of the inhibition
by its self-interaction. Since the self-interaction spike is then
missing at time $\theta+2T$, a spike will be emitted at that later
time and so on. So the spike sub-pattern of this neuron (consisting of all 
those spikes in the total pattern that are generated by neuron $l$) 
has period $2T$, and not $T$. 

However the spike sub-pattern of any neuron $l$ has to be periodic
even if it receives supra-threshold input. This can be seen as follows:
Due to the conventions above, a spike can only
be emitted when there is a discontinuity in the phase $\phi_{l}$
(after a supra-threshold excitation, the phase is always zero, after
a simultaneous reception and spiking it is always unequal to $\Theta_{l}$)
or if the neuron receives a supra-threshold input when its phase is
$\phi_{l}(\theta^{-})=0$.
 Since $\phi_l(t)$ is piecewise continuous,
in every (finite) time interval $[t,t+T)$
there are only finitely many discontinuities, as well as only finitely
many times with $\phi_{l}(\theta^{-})=0$ because the phase is monotonous
otherwise. Therefore, given a certain phase dynamics, spikes can be
emitted by the network only at finitely many times in any interval
$[t,t+T)$.
This implies that there are only finitely many combinations
of spikes which can be emitted by the network within a period $T$ of the
phases. Thus, after
a finite integer multiple of $T$, the spike patterns
have to recur. After this has happened, 
not only the phases but (because here we can choose $T$ to be an arbitrary
integer multiple of the phase period such that
$\tau_{lm}<T$ without loss of generality) also all spikes in transit are the same as
at some time before. Since at any time the state of the network is fixed by
the phases and the spikes in transit, the entire dynamics must repeat.
So, the pattern is periodic with some period $nT$,
$n\in\mathbb{{N}}$.

\begin{thm}\label{phaseperiod}
Let $S\subset\{1,...,N\}$ be the set of neurons that (i) do not receive
any supra-threshold excitations and (ii) are firing at least once
in the pattern. Then, the periodicity $T$ of the entire pattern is
sufficient for the periodicity of the \textbf{}phases \begin{equation}
\phi_{l}(t)=\phi_{l}(t+nT),\label{eq:P1}\end{equation}
 for all neurons $l\in S$, all $n\in\mathbb{Z}$ and all $t\in[0,T)$.
\end{thm}
We disprove the opposite: Suppose, for some $l\in S$ and some $t$,
$\phi_{l}(t)>\phi_{l}(t+T)$. Then this inequality remains true for
all future times $t$. First, it remains true during free time evolution.
Because the inputs are identical for every period and because the
$H_{\varepsilon}^{(l)}(\phi)$ are strictly monotonically increasing
as function of $\phi$, it remains true also after arbitrarily many
interactions. Therefore, denoting the next firing time of neuron $l$
after time $t$ by $t_{j}$, we conclude that $1=\phi_{l}(t_{j}^{-})>\phi_{l}((t_{j}+T)^{-})$,
violating the pattern's periodicity. An analogous argument shows that
if $\phi_{l}(t)<\phi_{l}(t+T)$ for some $t$, the pattern would not
be periodic either. Therefore, if the pattern is periodic, the phases
of neurons $l\in S$ are also periodic and 
the phases have the period of the pattern.

As direct consequence from Theorems \ref{pattperiod} and \ref{phaseperiod} we note the important special case $S=\{1,...,N\}$.

\begin{cor}
If all neurons in the network receive only subthreshold input and
are firing at least once in a pattern, periodicity of the entire pattern
is equivalent to the periodicity of the phase dynamics and the periods are
equal.
\end{cor}
\begin{rem}
If a neuron that (i) receives one or more supra-threshold inputs or
(ii) is silenced (i.e. has no firing time in the pattern) has
non-periodic phase dynamics, its spike sub-pattern can still be periodic. 
\end{rem}
(i) If a neuron $l$ receives a supra-threshold input, a small initial
deviation from the periodic phase dynamics that occurs sufficiently
briefly before the input, will only change the phase $\phi_{l}$ of
that neuron but not its next spike time as long as the input remains
supra-threshold. Since the dynamics continues without deviations with
respect to the periodic phase dynamics, all future events will also
take place at the predefined times. Thus there are initial conditions
such that the phase dynamics is not entirely periodic but the spike
pattern is. (ii) A sufficiently small initial deviation from the periodic
phase dynamics that occurs at a silenced neuron can decay without
making the neuron \textbf{}fire such \textbf{}that the spike pattern
stays periodic as without the deviation, although the phase of the
silenced neuron is not periodic.

For simplicity, we impose in the following that the phase dynamics
of all neurons, including those neurons that are silent (i.e. never
send a spike) and those that receive supra-threshold inputs, are periodic
with period $T$. We consider $\phi_{l}(t)$ for $t\in[0,T)$ with
periodic boundary conditions. All times are measured modulo $T$ and
spike time labels $j$ are reduced to $\{1,...,M\}$ by subtracting
a suitable integer multiple of $M$.

\subsection{Parameterizing all admissible network designs\label{sub:Parameterizing-all-admissible}}

In this subsection we are working towards an analytical restriction
of the set of all admissible networks for a given spike pattern. We provide a 
method of indexing all spike reception times, and of ordering them in time.The input coupling strengths are indexed accordingly.
Based on this scheme, we derive conditions ensuring the sending of a spike at the pre-defined spike times, periodicity of the phase dynamics, and quiescence (non-spiking) of the neurons between their desired spike times. 
A main result of the paper, Theorem \ref{thm:main}, provides a system of
restrictions on the coupling strengths, which separate into disjoint
constraints for the couplings onto each neuron, 
cf.\ Remark \ref{rem:constraintsSeparate}. 

Let
$\theta_{l,j}:=t_{j}+\tau_{ls_{j}}$ be the time when neuron $l$ receives
the spike labeled $j$ from neuron $s_{j}$. Then, for inhomogeneous
delay distribution the $\theta_{l,j}$ might not be ordered in $j$.
Therefore, we define a permutation $\sigma_{l}:\{1,...,M\}\rightarrow\{1,...,M\}$
of the indices of spikes received by neuron $l$, such that \begin{equation}
\bar{\theta}_{l,j}:=\theta_{l,\sigma_{l}(j)}\label{eq:OrderedArrivalTimes}\end{equation}
 is ordered, i.e. $\bar{\theta}_{l,j}\geq\bar{\theta}_{l,i}$ if $j>i$.
If multiple spikes are received at one time, $\sigma_{l}$ is not unique.
This, however, has no consequence for the collective dynamics because
all the associated spike receptions are treated as one according to
(\ref{eq:WellDefined}).

If neuron $l$ receives multiple, say $p(l,j)$ spikes at time $\bar{\theta}_{l,j}$,
we only consider the lowest of all indices $j'$ with reception time
$\bar{\theta}_{l,j'}=\bar{\theta}_{l,j}$. If neuron $l$ receives
spikes at $M_{l}$ \emph{different} times, we denote the smallest
index of each reception time by $j_{1}(l),...,j_{M_{l}}(l)$ such
that \begin{equation}
j_{n}(l):=j_{n-1}(l)+p(l,j_{n-1}(l)).\label{eq:SmallestIndex}\end{equation}
 for $n\in\{2,\ldots,M_{l}\}$. Here $j_{1}(l)=1.$ The first set
of equal reception times starts with index $j_{1}(l)=1$ and contains
$p(l,1)$ spikes. Therefore, the second set of equal reception times
has first index $j_{2}(l)=p(l,1)+1=p(l,j_{1}(l))+j_{1}(l)$ and contains
$p(l,j_{2}(l))$ spikes. This way all indices are defined recursively. 

\begin{figure}
\begin{center}\includegraphics[%
  width=90mm]{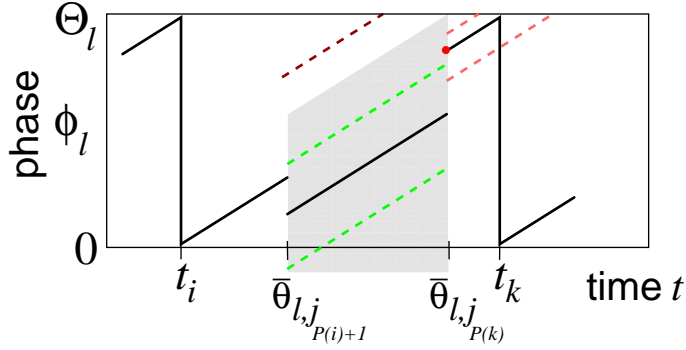}\end{center}
\caption{(color) Restriction of a neuron's dynamics between its firing events,
cf. (\ref{eq:F}) and (\ref{eq:S}). In this example, two spikes arrive
between the firing times $t_{i}$ and $t_{k}$ of neuron $l$. The
solid line indicates one possible time evolution of the phase $\phi_{l}(t)$.
Between the firing times, $\phi_{l}(t)$ may follow any path within
the possibly semi-infinite polygon (gray shaded; green dashed lines show other possible trajectories).
A too large phase at $\bar{\theta}_{l,j_{P(i)+1}}$ contradicts (\ref{eq:S})
and will lead to early firing (dark red dashed line). The phase at
$\bar{\theta}_{l,j_{P(k)}}$ is fixed (red dot). Any other phase inconsistent
with the equality in (\ref{eq:F}) would lead to a firing time earlier
or later than predefined (light red dashed lines). \label{cap:PatternN}}
\end{figure}

To keep the notation concise, we skip the argument $l$ in the following
(where it is clear) as the argument or index of some quantity which
is itself a further index or a subindex, e.g., of $\bar{\theta}_{l}$
or $\eps_{l}$. For instance, we abbreviate $\bar{\theta}_{l,j_{i}(l)}$
by $\bar{\theta}_{l,j_{i}}$ and $p(l,j_{k}(l))$ by $p(j_{k})$ where
appropriate. Furthermore, indices denoting different spike receptions
of neuron $l$ are reduced to $\{1,...,M_{l}\}$ by subtracting a
suitable multiple of $M_{l}$. We define $\PP_{l}(i)\in\{1,...,M_{l}\}$
(cf. also Fig. \ref{cap:PatternN})
as the index of the last reception time for
neuron $l$ before its firing time $t_i$,
\begin{equation}
\PP_{l}(i):=\text{argmin}\{ t_{i}-\bar{\theta}_{l,j_{k}}\,|\,
k\in\{1,...,M_{l}\},t_{i}-\bar{\theta}_{l,j_{k}}>0\}.\label{eq:LastSpikeBeforeFiring}\end{equation}
If there are no simultaneous spikes received
by neuron $l$ and if there is no spike received at the firing time
$t_{i}$ itself, $\PP_{l}(i)$ is given by \textbf{}\begin{equation}
\PP_{l}(i)=\text{argmin}\{ t_{i}-\bar{\theta}_{l,j}\,|\, j\in\{1,...,M\}\}.\label{eq:LastOneSpikeBeforeFiring}\end{equation}
In the following, if two or more reception times are equal, we will
select the smallest index and restrict the dynamics only \emph{once},
using Eqns.\ (\ref{eq:Hcomposition}),(\ref{eq:WellDefined}) and the definition of $j_{i}(l)$ above.
Only the total action of all spikes received by a neuron $l$ at a
particular $\bar{\theta}_{l,j_{i}}$ will be restricted, by a single condition.
We therefore define the sum of the coupling strengths of all spikes
received by neuron $l$ at time $\bar{\theta}_{l,j_{i}}$ as \begin{equation}
\bar{\varepsilon}_{l,i}=\varepsilon_{ls_{\sigma(j_{i})}}+...+\varepsilon_{ls_{\sigma(j_{i}+p(j_{i})-1)}}.\label{eq:epsTotal}\end{equation}
 Indeed, $\sigma_{l}(j_{i}(l)+k)$, $k\in\{0,...,p(l,j_{i}(l))-1\}$,
are the indices of the $p(l,j_{i}(l))$ different spikes received
by neuron $l$ at the $i$th reception time $\bar{\theta}_{l,j_{i}}$,
$i\in\{1,...,M_{l}\}$. If neuron $l$ receives all spikes at different
times, we have $\bar{\varepsilon}_{l,i}=\varepsilon_{ls_{\sigma(i)}}$.
Let \begin{equation}
\Delta_{l,i}=\bar{\theta}_{l,j_{i+1}}-\bar{\theta}_{l,j_{i}}\label{eq:Deltali}\end{equation}
 be the time differences between two successive different reception
times, where $i+1$ has to be reduced to $\{1,...,M_{l}\}$ by subtracting
a suitable integer multiple of $M_{l}$. We now rewrite Eqns.~(\ref{eq:F1})
and (\ref{eq:S1}) for neuron $l$ as a set of conditions on the phases
$\phi_{l}(\bar{\theta}_{l,j_{i}})$ at the different spike reception
times $\bar{\theta}_{l,j_{i}}$ in terms of the firing times $t_{i_{k}}$
of that neuron and the spike reception times $\bar{\theta}_{l,j_{i'}}$,
$i'\in\{1,...,M_{l}\}$. 

If the given pattern does not imply the reception of a spike precisely
at the firing time $t_{i_{k}}$ (together with the firing times and
the delays also the reception times are fixed), this results in\begin{align}
\phi_{l}(\bar{\theta}_{l,j_{\PP(i_{k})}})= & \Theta_{l}-(t_{i_{k}}\mT\bar{\theta}_{l,j_{\PP(i_{k})}}),\label{eq:F}\\
\phi_{l}(\bar{\theta}_{l,j_{i}})< & \Theta_{l}-\Delta{}_{l,i},\label{eq:S}\end{align}
 where $k\in\{1,...,K(l)\}$ and $i\in\{1,...,M_{l}\}\backslash\{\PP(i_{k})|k\in\{1,...,K(l)\}\}$.
We note that, by definition (\ref{eq:LastSpikeBeforeFiring}), there
is no input to neuron $l$ between the spike(s) received at $\bar{\theta}_{l,j_{\PP(i_{k})}}$
and the neuron's next firing time $t_{i_{k}}$. 

The firing time condition (\ref{eq:F}) states that the neuron at
time $\bar{\theta}_{l,j_{\PP(i_{k})}}$ is as far away from its threshold
$\Theta_{l}$ as it needs to be in order to exactly evolve there freely
in the remaining time $t_{i_{k}}\mT\bar{\theta}_{l,j_{\PP(i_{k})}}$.
The inequalities (\ref{eq:S}) guarantee that the neuron does not
spike between the firing times determined by the predefined pattern:
They ensure that neuron $l$ is far enough from its threshold at all
other spike reception times and is not firing at any time that is
not in the desired pattern, $t\neq t_{i_{k}}$. 

Above, we had fixed the convention, that if a spike is received by
a neuron
when it is just about to fire, the spike received is processed
after the sending of the new spike. If we had used the convention
that first the received spike is considered, the {}``$<$'' in inequality
(\ref{eq:S}) would have been replaced by a {}``$\leq$''. Here
equality, $\phi_{l}(\bar{\theta}_{l,j_{i}})=\Theta_{l}-\Delta{}_{l,i}$,
means that the neuron approaches the threshold at $\bar{\theta}_{l,j_{i+1}}^{-}$,
i.e. $\phi_{l}(\bar{\theta}_{l,j_{i+1}}^{-})=\Theta_{l},$ but since
the received spike is processed first, an untimely spike can be prevented
by an inhibitory input.

If there is one or several spikes received precisely at a predefined
firing time $t_{i_{k}}$, supra-threshold excitation can be used to
realize the pattern. To account for this, the firing time condition
(\ref{eq:F}) and the silence condition (\ref{eq:S}) with
$i=P_l(i_k)+1$ have to be replaced by the conditions
\begin{align}
\phi_{l}(\bar{\theta}_{l,j_{\PP(i_{k})}}) & <\Theta_{l}-(t_{i_{k}}\mT\bar{\theta}_{l,j_{\PP(i_{k})}}),\label{eq:FST1}\\
U_{l}(\phi_{l}(t_{i_{k}}^{-}))+\bar{\eps}_{l,\PP(i_{k})+1} & \geq
U_{l}(\Theta_{l}).\label{eq:FST2}
\end{align}
Here, the strict inequality (\ref{eq:FST1}) prevents untimely spiking
(cf. the dark red dashed line in Fig.~\ref{cap:PatternN}) and guarantees
that the neuron does not reach the threshold by its intrinsic
dynamics. The second, inequality (\ref{eq:FST2}), ensures the spiking
at $t_{i_{k}}$. However, (\ref{eq:FST2}) is not an inequality on
the phases depending at the reception times only, but involves the
total coupling of the incoming spikes. We note that expression (\ref{eq:FST1})
with an equal sign, ``$=$'', describes the case that the neuron spikes
without supra-threshold excitation, because due to our above convention,
the firing is treated before the spike reception. Then, inequality
(\ref{eq:FST2}) is obsolete. So Eq.~(\ref{eq:F}) is the appropriate
spike time condition also if spikes are received by neuron $l$ when
it just reaches threshold. Now, there are two cases possible (i) the
spikes do not cause a supra-threshold excitation $U_{l}(0)+\bar{\eps}_{l,\PP(i_{k})+1}<U_{l}(\Theta_{l})$
from the reset phase of the neuron or (ii) they cause a supra-threshold
excitation, $U_{l}(0)+\bar{\eps}_{l,\PP(i_{k})+1}\geq U_{l}(\Theta_{l})$.
In the first case, $\phi_{l}(t_{i_{k}})=\phi_{l}(\bar{\theta}_{l,j_{\PP(i_{k})+1}})=H^{(l)}_{\bar{\eps}_{l,\PP(i_{k})+1}}(0),$
in the second $\phi_{l}(t_{i_{k}})=\phi_{l}(\bar{\theta}_{l,j_{\PP(i_{k})+1}})=0$.
In the first case, the silence condition (\ref{eq:S}) with $i=\PP(i_{k})+1$
applies such that this case does not need a special treatment, in the second,
we have the inequality $\bar{\eps}_{l,\PP(i_{k})+1}\geq U_{l}(\Theta_{l})$ instead.

Specifying conditions on the phases at these ordered and clustered
(simultaneous) spike reception times is equivalent to specifying the
phases at the
unordered and unclustered times because $\phi_{l}(\theta_{l,i})=\phi_{l}(\theta_{l,j})$
if $\theta_{l,i}=\theta_{l,j}$. 

If there are no simultaneous events, the strengths of coupling onto
a particular neuron $l$, $\eps_{ll'}$, $l'\in\{1,\ldots,N\}$, are
restricted by $K(l)$ nonlinear equations and $M-K(l)$ inequalities
originating from (\ref{eq:F}) and (\ref{eq:S}). All the coupling
strengths in the network realizing a given pattern are thus restricted
by a system of $\sum_{l=1}^{N}K(l)=M$ nonlinear equations and $\sum_{l=1}^{N}(M-K(l))=(N-1)M$
inequalities.

\begin{rem}
The constraints (equations and inequalities) restricting the coupling
strengths of the network (to be consistent with a predefined pattern)
separate into disjoint constraints for the couplings onto each individual
neuron. \label{rem:constraintsSeparate}
\end{rem}
In the presence of simultaneous events, for each neuron there
are $M_{l}-K(l)+S(l)$ inequalities originating from (\ref{eq:FST1}),
(\ref{eq:FST2}) and (\ref{eq:S}), (where $S(l)$ is the number of
supra-threshold excitations, not counting the ones where the spike
is omitted) and $K(l)-S(l)$ equations originating from the spikings
described by (\ref{eq:F}). We see that simultaneous receptions decrease
the number of constraints. Again, these constraints separate (remark
\ref{rem:constraintsSeparate}). This property is due to the fact that
the pattern is fixed; it turns out (see below) that because of this 
separation, it is easier to find a solution for the coupling strengths that satisfy these constraints.

Fig.~\ref{cap:PatternN} illustrates the constraints. After a firing
of neuron $l$ at time $t_{i}$ where its phase is zero, conditions
(\ref{eq:F}) and (\ref{eq:S}) impose restrictions on the phases
at the spike reception times while the time evolution proceeds towards
the subsequent firing time $t_{k}$ of neuron $l$. 

If we now compute explicitely the dynamics of neuron $l$ between
two successive firing times $t_{i}$ and $t_{k}$ and evaluate the
dynamics at the times occurring in (\ref{eq:F}) and (\ref{eq:S}),
we obtain 

\begin{equation}
\begin{array}{rl}
H_{\bar{\eps}_{l,\PP(i)\pM1}}^{(l)}(\bar{\theta}_{l,j_{\PP(i)\pM1}}-t_{i}) & <\Theta_{l}-\Delta{}_{l,\PP(i)\pM1}\,,\\
H_{\bar{\eps}_{l,\PP(i)\pM2}}^{(l)}(H_{\bar{\eps}_{l,\PP(i)\pM1}}^{(l)}(\bar{\theta}_{l,j_{\PP(i)\pM1}}-t_{i})+\Delta{}_{l,\PP(i)\pM1}) & <\Theta_{l}-\Delta{}_{l,\PP(i)\pM2}\,,\\
 & \,\,\vdots\\
H_{\bar{\eps}_{l,\PP(k)}}^{(l)}(...H_{\bar{\eps}_{l,\PP(i)\pM2}}^{(l)}(H_{\bar{\eps}_{l,\PP(i)\pM1}}^{(l)}(\bar{\theta}_{l,j_{\PP(i)\pM1}}-t_{i})+\Delta{}_{l,\PP(i)\pM1})\\
\ldots+\Delta{}_{l,\textrm{}\PP(k)\mM1}) & =\Theta_{l}-(t_{k}-\bar{\theta}_{l,j_{\PP(k)}})\end{array}\label{eq:F2A}\end{equation}

in the case of no spike reception at time $t_{i}$ and no supra-threshold
excitation that generates the spike at $t_{k}$. 

Now we consider the case that there was a spike reception at time
$t_{i}$. If a supra-threshold spike generated the spike time $t_{i}$
from a phase $\phi_{l}(t_{i}^{-})<\Theta_{l}$ and the intrinsic dynamics
generates the spike at $t_{k}$, the set of equations and inequalities
reads \begin{equation}
\begin{array}{rl}
H_{\bar{\eps}_{l,\PP(i)\pM2}}^{(l)}(\Delta{}_{l,\PP(i)\pM1}) & <\Theta_{l}-\Delta{}_{l,\PP(i)\pM2}\,,\\
 & \,\,\vdots\\
H_{\bar{\eps}_{l,\PP(k)}}^{(l)}(...H_{\bar{\eps}_{l,\PP(i)\pM2}}^{(l)}(\Delta{}_{l,\PP(i)\pM1})\ldots+\Delta{}_{l,\PP(k)\mM1}) & =\Theta_{l}-(t_{k}-\bar{\theta}_{l,j_{\PP(k)}}).\end{array}\label{eq:F2B}\end{equation}

Alternatively, at $t_{i}$, the threshold can be reached by the intrinsic dynamics
$\phi_{l}(t_{i}^{-})=\Theta_{l}$ although a spike is arriving. Here
we have to consider two different cases: (i) $U_{l}(0)+\bar{\eps}_{l,\PP(i)\pM1}<U_{l}(\Theta_{l})$,
i.e. the spike is subthreshold. This is just a special case of (\ref{eq:F2A})
with $\bar{\theta}_{l,j_{\PP(i)\pM1}}-t_{i}=0$. (ii) $U_{l}(0)+\bar{\eps}_{l,\PP(i)\pM1}\geq U_{l}(\Theta_{l})$,
i.e. the spike is supra-threshold. In this case, we fixed the convention
that the second spike is omitted and the neuron is reset to zero;
therefore system (\ref{eq:F2B}) is supplemented with the
condition
\begin{equation}
\bar{\eps}_{l,\PP(i)\pM1}\geq U_{l}(\Theta_{l})\label{eq:F2BAddCond}\end{equation}
on $\bar{\eps}_{l,\PP(i)\pM1}$. 

The above equations also cover the case that a spike is received by
neuron $l$ at the spike time $t_{k}$ when neuron $l$ already reached
$\Theta_{l}$, i.e. $\bar{\theta}_{l,j_{\PP(k)\pM1}}=t_{k}$.
However, also supra-threshold excitation can then also
be used to generate the spike $t_{k}$. Then, if no spike is received
at $t_{i}$, or if a spike is received when the threshold is already reached
and no supra-threshold excitation takes place, the couplings are restricted
by (\ref{eq:F2A}) where the last equation has to be replaced by the
inequalities

\begin{align}
H_{\bar{\eps}_{l,\PP(k)}}^{(l)}(...H_{\bar{\eps}_{l,\PP(i)\pM2}}^{(l)}(H_{\bar{\eps}_{l,\PP(i)\pM1}}^{(l)}(\bar{\theta}_{l,j_{\PP(i)\pM1}}-t_{i})\nonumber \\
+\Delta{}_{l,\PP(i)\pM1})\ldots+\Delta{}_{l,\PP(k)\mM1}) & <\Theta_{l}-(t_{k}-\bar{\theta}_{l,j_{\PP(k)}}),\nonumber \\
U_{l}(H_{\bar{\eps}_{l,\PP(k)}}^{(l)}(...H_{\bar{\eps}_{l,\PP(i)\pM2}}^{(l)}(H_{\bar{\eps}_{l,\PP(i)\pM1}}^{(l)}(\bar{\theta}_{l,j_{\PP(i)\pM1}}-t_{i})\nonumber \\
+\Delta{}_{l,\PP(i)\pM1})\ldots+\Delta{}_{l,\PP(k)\mM1})+\Delta{}_{l,\PP(k)})+\bar{\eps}_{l,\PP(k)+1} & \geq U_{l}(\Theta_{l}).\label{eq:F2C}\end{align}

If supra-threshold excitation occurred at time $t_{i}$ and supra-threshold input generated the spike at $t_{k}$, the couplings are restricted by (\ref{eq:F2B}) (possibly completed by (\ref{eq:F2BAddCond})), where the last equation
has to be replaced by the inequalities
\begin{align}
H_{\bar{\eps}_{l,\PP(k)}}^{(l)}(...H_{\bar{\eps}_{l,\PP(i)\pM2}}^{(l)}(\Delta{}_{l,\PP(i)\pM1})\nonumber \\
\ldots+\Delta{}_{l,\PP(k)\mM1}) & <\Theta_{l}-(t_{k}-\bar{\theta}_{l,j_{\PP(k)}}),\nonumber \\
U_{l}(H_{\bar{\eps}_{l,\PP(k)}}^{(l)}(...H_{\bar{\eps}_{l,\PP(i)\pM2}}^{(l)}(\Delta{}_{l,\PP(i)\pM1})\nonumber \\
\ldots+\Delta{}_{l,\PP(k)\mM1})+\Delta{}_{l,\PP(k)})+\bar{\eps}_{l,\PP(k)+1} & \geq U_{l}(\Theta_{l}).\label{eq:F2D}\end{align}

We have thus shown:

\begin{thm}
The set of solutions to the systems (\ref{eq:F2A})--(\ref{eq:F2D})
\textbf{}for all $K(l)$ pairs of subsequent firing times $(t_{i},t_{k})$,
where $i=i_{n}$, $k=i_{n+1}$, $n\in\{1,\ldots,K(l)\}$, provides
the set of all admissible coupling strengths $\eps_{ll'}$, $l'\in\{1,\ldots,N\}$,
of incoming connections to neuron $l$. 
\label{thm:main}
\end{thm}
\begin{cor}
Solutions to systems analogous to (\ref{eq:F2A})--(\ref{eq:F2D})
for all neurons $l\in\{1,\ldots,N\}$ define all coupling strengths
of an admissible network.
\begin{figure}
\begin{center}\includegraphics[%
  width=88mm]{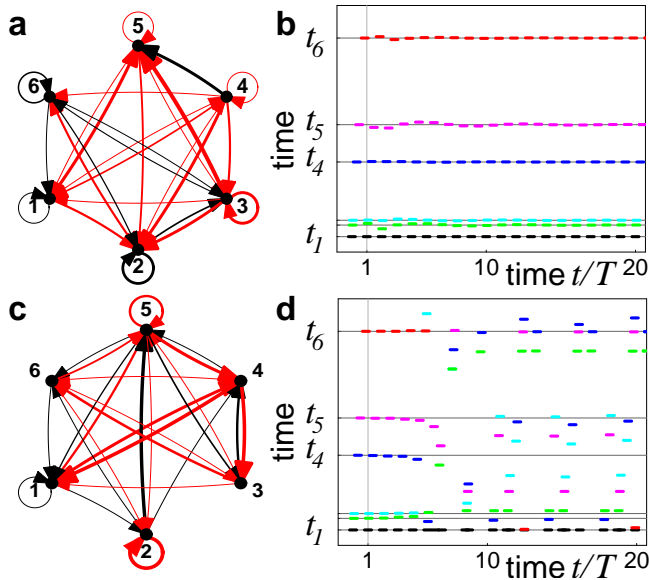}\end{center}

\caption{(color online) Two different networks (a), (c) realize the same predefined
pattern ((b), (d) grey lines). The networks consist of six identical
leaky integrate-and-fire neurons with $I_{m}=1.2$, $\gamma_{m}=1$,
$\Theta_{m}=1$. The networks are realizations of random graphs where
each coupling is present with probability $p=0.8$; the coupling delay
is $\tau_{ml}=0.125$. A small random perturbation is applied at the
beginning of the second period. The network dynamics (spike times
relative to the spikes of neuron $l=1$, color coded for each neuron),
found by exact numerical integration \cite{TWG03} shows that in network
(a) the pattern is stable and thus regained after a few periods (b);
in network (c) the pattern is unstable and eventually another
pattern is assumed (d). Reproduced from Ref.~\cite{MT05}. \label{cap:Single-periodic-patterns}}
\end{figure}

\end{cor}
Often (\ref{eq:F2A})--(\ref{eq:F2D}) are under-determined systems such that
many solutions exist, implying that many different networks realize the same
predefined pattern, cf. Fig.\ \ref{cap:Single-periodic-patterns}. This is
illustrated in more detail in the next section. Roughly speaking, in the
absence of supra-threshold excitation, the time of each spike of each neuron
provides one ``hard'' (equality) constraint on the in general N-dimensional
set of input coupling strengths of that neuron. The silence conditions provide
``soft'' (inequality) constraints, often not lowering the dimensionality of
the solution space of coupling strengths. Intuitively a hard restriction can
be understood by considering a simple example: Consider a network of $N=3$
neurons. If one neuron $m$ receives two spikes in a fixed time interval in
which it does not send a spike itself, the coupling strengths of these spikes
are arbitrary as long as their \emph{total impact} on the neuron's phase
$\phi_m$ (advancing or retarding) is the same, cf. also Fig.
\ref{cap:PatternN}. This provides one, and not two, hard restrictions to the
set of input coupling strengths to neuron $m$.

In the case of leaky integrate-and-fire or Mirollo-Strogatz neurons,
a solution of (\ref{eq:F2A})--(\ref{eq:F2D}),
if one exists, can be found in a simple way, because the system is
then reducible to be linear in the coupling strengths or polynomial in its
exponentials, respectively.

\begin{rem}
There are patterns for which the systems (\ref{eq:F2A})--(\ref{eq:F2D}),
with predefined neuron properties and predefined delay distribution,
do not have a solution.
\end{rem}
This means that if the delays and neural parameters are specified,
no network, independent of how the coupling strengths are chosen,
exhibits that predefined pattern. This can already be observed from
a simple example: consider a non-degenerate pattern where neuron $l$ 
sends  three  successive spikes and between each two successive of these 
spike times there is precisely one spike received, each sent by the same 
neuron $m$. Then, the coupling strength $\eps_{lm}$ is fixed (by the
firing time condition to which (\ref{eq:F2A}) reduces) to ensure the correct
time of the second spike of neuron $l$  and cannot
be modified to ensure the third one. So, if the interval between the second
and third spike time does not by coincidence match the one determined by the input, the pattern will not be realizable by any
network. Other, more complicated examples follow immediately. 

This implies that certain predefined patterns may not be realizable
in any network, no matter how its neurons are interconnected. We note
that if we allow the neural parameters and delay times to vary as
well, the system again might have a solution. \textbf{}

\subsection{Explicit analytical parameterization}\label{sub:ExplAnaPara}

In this sub-section, we will show that an entire class of patterns
can, under few weak requirements always be realized by a (typically
multi-dimensional) family of networks. This class consists of simple
periodic patterns, in which every neuron fires exactly once before
the pattern repeats. For a simple periodic pattern, we label, without
loss of generality, the neuron firing at time $t_{l}$ by $l$, i.e.\ $s_{l}=l$
for $l\in\{1,...,M = N\}$. Accordingly we have $\theta_{l,m}=t_{m}+\tau_{lm}$.
The time differences between two successive spike times of the same
neuron equal the period of the simple periodic pattern. Thus, for
each neuron $l$ the reception times of spikes from all neurons of
the network are guaranteed to lie between two successive firings of
neuron $l$.
We note again, that due to the periodicity of the pattern, we can assume
without loss of generality that the delay times are smaller than the
patterns period; otherwise, we take them modulo $T$ without changing
the invariant dynamics. In the following, we require that two simple
criteria are met. 

\begin{crit}
For each neuron its self-interaction delay is smaller than its free period,
i.e. $\tau_{ll}<T_{0,l}$ for $l\in\{1,\ldots,N\}$.\label{cri:MaxDelays}
\end{crit}

This criterion ensures that the spike time of each neuron can be modified,
at least by the self-coupling. If, as we assume throughout the manuscript
(see section \ref{sec:NetworkDesign}),
a neuron $l$ firing only once in the period (here at $t_l$) receives at least
one spike in the interval $(t_l,t_l+\Theta_l)$ (or, if $\Theta_l \geq T$ in $(t_l,t_l+T)$),
this criterion is not necessary to hold for Theorem
\ref{thm:SimplePeriodicPatternManifold} below; Theorem
\ref{thm:SimplePeriodicPatternManifold}
 holds for any presynaptic neuron sending the spike modifying the spike time
 (Criterion \ref{cri:ThresholdReachable} appropriately modified).

\begin{crit}
The threshold minus a possible lower bound of the phase plus the self-interaction
delay for each neuron $l$ is larger than the pattern's period, $\Theta_{l}-B_{l}+\tau_{ll}>T$.
\label{cri:ThresholdReachable} 
\end{crit}

This second condition is obsolete if there is no lower bound of the
phase, as e.g.~for leaky integrate-and-fire neurons. 

Given these weak constraints, the following statement holds. 

\begin{thm}
For simple periodic patterns, if conditions (\ref{cri:MaxDelays})
and (\ref{cri:ThresholdReachable}) are satisfied, solutions to (\ref{eq:F2A})
exist and the set of admissible networks contains an $N(N-1)$ dimensional submanifold in the space
of coupling strengths. \label{thm:SimplePeriodicPatternManifold}
\end{thm}
This means that all simple periodic patterns are typically realizable
by a high-dimensional family of networks.

We first show that one solution exists, then state another Theorem,
which explicitly shows that the solution space contains an $N(N-1)$-dimensional
submanifold. 

We explicitly construct a trivial solution, where only self-interaction
is present, while all the other coupling strengths are zero. We consider the
one neuron system consisting of neuron $l$. Because of $\phi_{l}(t_{l})=0$
and condition (\ref{cri:MaxDelays}) at the reception time of the
spike from neuron $l$ to itself, $\phi_{l}((t_{l}+\tau_{ll})^{-})=\tau_{ll}$
holds. At time $t_{l}+\tau_{ll}$ the neuron's phase is set to $\phi_{l}(t_{l}+\tau_{ll})=\Theta_{l}-(T-\tau_{ll})<\Theta_{l}$
by choosing the coupling strength $\eps_{ll}=H_{\phi_{l}(t_{l}+\tau_{ll})}^{(l)-1}(\phi_{l}((t_{l}+\tau_{ll})^{-}))$.
Here, $H_{\psi}^{(l)-1}(\phi)=U_{l}(\psi)-U_{l}(\phi)$ is the inverse
of $H_{\eps}^{(l)}(\phi)$ with respect to $\eps$, which exists for
any $\psi$ and $\phi$ in the domain of $U_{l}$. Indeed, $0\leq\phi_{l}((t_{l}+\tau_{ll})^{-})<\Theta_{l}$
is in the domain of $U_{l}$ as well as $\phi_{l}(t_{l}+\tau_{ll})$.
The latter is true, even if a lower bound is present, because $\phi_{l}(t_{l}+\tau_{ll})=\Theta_{l}-(T-\tau_{ll})>B_{l}$
due to condition \ref{cri:ThresholdReachable}. Now, since no further
spike is received, the condition Eq.~(\ref{eq:F}) for the spike
sending time is satisfied and the next spiking will take place at
$t_{l}+T$. Since there are no further spike receptions there are
no silence conditions (\ref{eq:S}) to be satisfied. All neurons taken
together as a network without couplings between different neurons
the pattern is invariant. We now set out to parameterize the entire
nonempty class of solutions realizing the given pattern. Indeed, for
simple periodic patterns this can be done analytically:

\begin{thm}
For any simple periodic pattern, the set of all networks satisfying
the systems (\ref{eq:F2A}-\ref{eq:F2D}) can be explicitly parameterized.\label{thm:SimplePeriodParam} 
\end{thm}
The parameterization for each neuron $l\in\{1,\ldots,N\}$ is given
as follows

(i) in the case $\theta_{l,j}\neq t_{l}$ for all $j\in\{1,...,N\}$,

\begin{align}
\bar{\eps}_{l,\PP(l)\pM1}= & H_{\phi_{l}(\bar{\theta}_{l,j_{\PP(l)\pM1}})}^{(l)-1}(\bar{\theta}_{l,j_{\PP(l)\pM1}}\mT t_{l}),\nonumber \\
\bar{\eps}_{l,\PP(l)\pM k}= & H_{\phi_{l}(\bar{\theta}_{l,j_{\PP(l)\pM k}})}^{(l)-1}(\phi_{l}(\bar{\theta}_{l,j_{\PP(l)\pM k-1}})+\Delta_{l,\PP(l)\pM k\mM1}),\nonumber \\
\bar{\eps}_{l,\PP(l)}= & H_{\Theta_{l}-(t_{l}\mT\bar{\theta}_{l,j_{\PP(l)}})}^{(l)-1}(\phi_{l}(\bar{\theta}_{l,j_{\PP(l)\mM1}})+\Delta_{l,\PP(l)\mM1}),\label{eq:epsi1}\end{align}

where $k\in\{2,...,M_{l}-1\}$ and the neurons' phases $\phi_{l}(\bar{\theta}_{l,j_{i}})$,
$i\in\{1,...,M_{l}\}\backslash\{ P_{l}(l)\}$ at the spike reception
times are the parameters that are subject to the restrictions (\ref{eq:S}).
These equations also hold with $\bar{\theta}_{l,j_{\PP(l)\pM1}}\mT t_{l}=0$
if there is a spike reception at $t_{l}$ but no supra-threshold excitation.

(ii) If there is a spike reception at $t_{l}$, neuron $l$ already reaches
threshold due to its intrinsic dynamics $\phi_l(t_l^-)=\Theta_l$, and there 
is supra-threshold excitation immediately after the reset, we have

\begin{align}
\bar{\eps}_{l,\PP(l)\pM1}\geq & U_{l}(\Theta_{l})-U_{l}(0),\nonumber \\
\bar{\eps}_{l,\PP(l)\pM2}= & H_{\phi_{l}(\bar{\theta}_{l,j_{\PP(l)\pM2}})}^{(l)-1}(\bar{\theta}{}_{l,j_{\PP(l)\pM2}}\mT t_{l}),\nonumber \\
\bar{\eps}_{l,\PP(l)\pM k}= & H_{\phi_{l}(\bar{\theta}_{l,j_{\PP(l)\pM k}})}^{(l)-1}(\phi_{l}(\bar{\theta}_{l,j_{\PP(l)\pM k-1}})+\Delta_{l,\PP(l)\pM k\mM1}),\nonumber \\
\bar{\eps}_{l,\PP(l)}= & H_{\Theta_{l}-(t_{l}\mT\bar{\theta}_{l,j_{\PP(l)}})}^{(l)-1}(\phi_{l}(\bar{\theta}_{l,j_{\PP(l)\mM1}})+\Delta_{l,\PP(l)\mM1}),\label{eq:epsi2}\end{align}

where $k\in\{3,...,M_{l}-1\}$. The parameters are the neurons' phases
$\phi_{l}(\bar{\theta}_{l,j_{i}})$,
$i\in\{1,...,M_{l}\}\backslash\{ P_{l}(l),P_{l}(l)+1\}$ at the spike
reception times that are subject to the restrictions
(\ref{eq:S}) and  $\bar{\eps}_{l,\PP(l)\pM1}$ which is bounded below by $\bar{\eps}_{l,\PP(l)\pM1}\geq U_{l}(\Theta_{l})$.

(iii) If there is a spike reception at $\theta_{l,j}=t_{l}$, and the spike at
$t_l$ is generated by
supra-threshold excitation:

\begin{align}
\bar{\eps}_{l,\PP(l)\pM2}= & H_{\phi_{l}(\bar{\theta}_{l,j_{\PP(l)\pM2}})}^{(l)-1}(\bar{\theta}_{l,j_{\PP(l)\pM2}}\mT t_{l}),\nonumber \\
\bar{\eps}_{l,\PP(l)\pM k}= & H_{\phi_{l}(\bar{\theta}_{l,j_{\PP(l)\pM k}})}^{(l)-1}(\phi_{l}(\bar{\theta}_{l,j_{\PP(l)\pM k-1}})+\Delta_{l,\PP(l)\pM k\mM1}),\nonumber \\
\bar{\eps}_{l,\PP(l)\pM1}\geq & U_{l}(\Theta_{l})-U_l(\phi_{l}(\bar{\theta}_{l,j_{\PP(l)}})+\Delta_{l,\PP(l)}),\label{eq:epsi3}\end{align}

where $k\in\{3,...,M_{l}\}$.
Here the parameters are the neurons' phases $\phi_{l}(\bar{\theta}_{l,j_i})$,
$i\in\{1,...,M_l\}\backslash\{ P_{l}(l)+1\}$ at the spike reception times
that are subject to the restrictions (\ref{eq:S}), (\ref{eq:FST1}) and
$\bar{\eps}_{l,\PP(l)\pM1}$, which 
is not parameterized but only bounded below by a function of $\phi_{l}(\bar{\theta}_{l,j_{\PP(l)}})$
unless we require that the spike precisely excites the neuron to the
threshold, i.e. the ``$=$'' in the last equation is valid. 

These relations follow directly from (\ref{eq:F2A}-\ref{eq:F2D})
by inversion and (\ref{eq:F}-\ref{eq:FST1}).

Since the
$\bar{\eps}_{l,i}$ are disjoint sums of couplings $\eps_{lj}$ , the
couplings towards neuron $l$ can be parameterized using the parameters for
$\bar{\eps}_{l,i}$ 
and $p(l,j_i)-1$ independent couplings per reception time $\bar{\theta}_{l,j_i}$.

We now demonstrate the second statement of Theorem \ref{thm:SimplePeriodicPatternManifold}.

In case (i) above, the
Jacobian of the couplings with respect to the phases can be directly seen
to have full rank $M_{l}-1$. Therefore, parameterization (\ref{eq:epsi1}) gives
an $M_{l}-1$-dimensional submanifold of the $M_{l}$-dimensional
space of $\bar{\eps}_{l,i}$. Since the $\bar{\eps}_{l,i}$ are just disjoint
sums of couplings $\eps_{lj}$, an $(N-1)$-dimensional submanifold
of networks realizing the pattern exists in $N$-dimensional $\eps_{lj}$-space,
$j\in\{1,\ldots,N\}$, $l$ fixed. We further know that the trivial
solution of uncoupled neurons with self-interaction constructed above is contained in case (i).
 Therefore, the set of parameters subject to the restrictions
(\ref{eq:S}) is nonempty. Since
it is open, there is an $(N-1)$-dimensional open set parameterizing
the submanifold. The product of these submanifolds of all couplings
is an $N(N-1)$-dimensional submanifold which is contained in the
set of solutions.

\subsection{A note on stability }

Is a pattern emerging in a heterogeneous network stable or unstable?
We numerically investigated patterns in a variety of networks and
found that in general the stability properties of a pattern depend
on the details of the network it is realized in, see Fig.\ \ref{cap:Single-periodic-patterns}
for an illustration. Depending on the network architecture, the same
pattern can be exponentially stable or unstable, or exhibit oscillatory
stable or unstable dynamics.

For any specific pattern in any specific network, the linear stability
properties can also be determined analytically, similar to the exact
perturbation analyses for much simpler dynamics in more homogeneous
networks \cite{TWG02a,TWG02b}. More generally, in every network of
neurons with congenerically curved rise functions and with purely
inhibitory (or purely excitatory) coupling, a nonlinear stability
analysis \cite{MT06} shows that the possible non-degenerate patterns
are either \emph{all} stable or \emph{all} unstable. For instance,
in purely inhibitory networks of neurons with rise functions of negative
curvature, such as standard leaky integrate-and-fire neurons, Eq.\ (\ref{eq:UIF})
with $\gamma>0$, every periodic non-degenerate spike pattern, no
matter how complicated, is stable.

If in the pattern, a neuron receives a spike when it was just about
to spike and the corresponding input coupling strength is not zero,
the pattern is super-unstable: an arbitrarily small perturbation in
the reception time can lead to a large change in the dynamics. These
cases, however, are very atypically in the sense that when randomly
drawing the delay times and the spike times in a pattern from a smooth
distribution the probability of occurrence of 
any simultaneous events, in particular those leading to this
super-instability, is zero. Simultaneous spikes sent and simultaneous spike received
by different neurons do not lead to a super-unstable pattern, because
the phase dynamics depends continuously on perturbations.

\section{Implementing additional requirements:\protect \\
Network Design on Predefined Connectivities\label{sec:AdditionalRequirements}}

\subsection{Can we require further system properties?}

As we have seen above, the systems of equations and inequalities (\ref{eq:F2A})--(\ref{eq:F2D})
\textbf{}defining the set of admissible networks is often underdetermined.
We can then require additional properties from the neurons and their
interactions. So far we assumed that neurons and delays were given
but arbitrary, but network coupling strengths, and therefore the connectivity,
were not restricted. 

Here we provide examples of how to require in advance additional features
that are controlled by the coupling strengths. A connection from a
neuron $l$ to $m$ can be absent (requiring the coupling strength
$\eps_{ml}=0$), taken to be inhibitory ($\eps_{ml}<0$) or excitatory
($\eps_{ml}>0$) or to lie within an interval; in particular, we can
specify inhibitory and excitatory subpopulations. 

Additional features entail additional conditions on the phases at
the spike reception times which can be exploited for network parameterization,
as we here demonstrate for simple periodic patterns, where we employ
the same conventions as in sub-section \ref{sub:ExplAnaPara}.

\emph{(i)} If the pattern is non-degenerate\emph{, exclusion of self-interaction}
is guaranteed by the conditions \begin{equation}
\phi_{l}(\theta_{l,l})=\tau_{ll}\label{eq:noselfWWnospike}\end{equation}
 if there is no spike-reception in $(t_{l},\theta_{l})$, and \begin{equation}
\phi_{l}(\theta_{l,l})-\phi_{l}(\theta_{l,\sigma(\sigma^{-1}(l)\mM1)})=\Delta_{l,\sigma^{-1}(l)\mM1}\label{eq:noselfWWspike}\end{equation}
 otherwise, typically reducing the dimension of the submanifold of possible
networks by $N$. 

\emph{(ii) Requiring} \emph{purely inhibitory networks}
leads to the accessibility conditions\begin{align}
\phi_{l}(\bar{\theta}_{l,j_{\PP(l)\pM1}})\leq & \bar{\theta}_{l,j_{\PP(l)\pM1}}\mT t_{l},\label{eq:purelyinhibitory1}\\
\phi_{l}(\bar{\theta}_{l,j_{i\pM1}})-\phi_{l}(\bar{\theta}_{l,j_{i}})\leq & \Delta_{l,i},\label{eq:purelyinhibitory2}\end{align}
 where $i\in\{1,...,M_{l}\}\backslash \{\PP_l(l)\}$. Since $\phi_{l}(\bar{\theta}_{l,j_{\PP(l)\pM1}}^{-})=\bar{\theta}_{l,j_{\PP(l)\pM1}}\mT t_{l}$,
the first inequality is equivalent to $\phi_{l}(\bar{\theta}_{l,j_{\PP(l)\pM1}})\leq\phi_{l}(\bar{\theta}_{l,j_{\PP(l)\pM1}}^{-})$.
This guarantees 
$\bar{\eps}_{l,\PP(l)\pM1}=H_{\phi_{l}(\bar{\theta}_{l,j_{\PP(l)\pM1}})}^{(l)-1}\left(\phi_{l}(\bar{\theta}_{l,j_{\PP(l)\pM1}}^{-})\right)=U_{l}(\phi_{l}(\bar{\theta}_{l,j_{\PP(l)\pM1}}))-U_{l}(\phi_{l}(\bar{\theta}_{l,j_{\PP(l)\pM1}}^{-}))\leq0,$
due to the monotonicity of $U_{l}$, such that the couplings summing up to
$\bar{\eps}_{l,\PP(l)\pM1}$ can be chosen to be inhibitory or zero. Analogously, the second inequality
ensures $\phi_{l}(\bar{\theta}_{l,j_{i}})\leq\phi_{l}(\bar{\theta}_{l,j_{i}}^{-}).$
We note that (\ref{eq:purelyinhibitory1}) also covers the case of
spikes received at time $t_{l}$. Since their action is inhibitory,
no supra-threshold excitation can occur and (\ref{eq:purelyinhibitory1})
yields $\phi_{l}(t_{l})=\phi_{l}(\bar{\theta}_{l,j_{\PP(l)\pM1}})\leq\bar{\theta}_{l,j_{\PP(l)\pM1}}\mT t_{l}=0$. 

To parameterize all networks we can therefore successively choose
$\phi_{l}(\bar{\theta}_{l,j_{\PP(l)\pM m}})$, $m\in\{1,...,M_l-1\}$,
starting with $m=1$. Inequalities (\ref{eq:purelyinhibitory1}) and
(\ref{eq:purelyinhibitory2}) hold with reversed relations for \emph{purely
excitatory coupling} if no supra-threshold excitation occurs. Otherwise,
they have to be replaced by

\begin{align}
\phi_{l}(\bar{\theta}_{l,j_{\PP(l)\pM2}})\geq & \bar{\theta}_{l,j_{\PP(l)\pM2}}\mT t_{l},\label{eq:purelyexcitatory1}\\
\phi_{l}(\bar{\theta}_{l,j_{i\pM1}})-\phi_{l}(\bar{\theta}_{l,j_{i}})\geq & \Delta_{l,i},\label{eq:purelyexcitatory2}\end{align}

where $i\in\{1,...,M_{l}\}\backslash\{ P_{l}(l),P_{l}(l)+1\}$. An
additional condition at time $t_{l}=\bar{\theta}_{l,j_{\PP(l)\pM1}}$
is not necessary, since the condition that the spike has a supra-threshold
action already ensures the excitatory coupling. In general, purely
inhibitory realizations can exist if the minimal inter-spike-interval
of each single neuron $l$ is larger than the neuron's free period,
i.e.\begin{equation}
\min\left\{ t_{i_{k+1}}-t_{i_{k}}|k\in\{1,...,K(l)\}\right\} \geq\Theta_{l},\label{eq:InhibCond}\end{equation}
 for all $l\in\{1,...,N\}$, where the index $k+1$ has to be reduced
to $\{1,...,K(l)\}$ subtracting a suitable multiple of $K(l)$. If
(\ref{eq:InhibCond}) is not satisfied, for some $k$, $\phi_{l}(t_{i_{k+1}}^{-})=\Theta_l$
is not reachable from $\phi_{l}(t_{i_{k}})=0$. For the same reason,
purely excitatory realizations can exist if\begin{equation}
\max\left\{ t_{i_{k+1}}-t_{i_{k}}|k\in\{1,...,K(l)\}\right\} \leq\Theta_{l}.\label{eq:ExzitCond}\end{equation}

In the case of simple periodic patterns, for purely inhibitory coupling
the inequalities (\ref{eq:InhibCond}) reduce to $T\geq\max_{m}\Theta_{m}$.
If even \begin{equation}
T>\max_{m}\Theta_{m}\label{eq:TlargerAllThres}\end{equation}

holds, the trivial solution is purely inhibitory with couplings $\eps_{ll}<0$.
Therefore, from Theorems \ref{thm:SimplePeriodicPatternManifold},
\ref{thm:SimplePeriodParam} and the corresponding proof, we conclude
that there is a submanifold of purely inhibitory networks in the
set of solutions. Analogously, if \begin{equation}
T<\min_{m}\Theta_{m},\label{eq:TsmallerAllThres}\end{equation}

there is a submanifold of purely excitatory networks in the set of
solutions.

\subsection{Very different connectivities, yet the same pattern }

Requiring certain connections to be absent is particularly interesting.
This just enters the restricting \textbf{}conditions
(\ref{eq:F2A}-\ref{eq:F2D}) as simple additional equalities $\eps_{ml}=0$
specifying that there is no connection from $l$ to $m$. 

By specifying absent connections we generally also specify which connections
are \emph{present} (except in cases where $\eps_{ml}=0$ by coincidence),
i.e. the \emph{connectivity} of the network. Though very simple to
implement, specifying the absence of connections is thus a very powerful
tool.

\begin{rem}
Absence of each of the $N^{2}$ connections $\eps_{ml}$, $m,l\in\{1,\ldots,N\},$
can be pre-specified independently.
\end{rem}
This means that we can typically \emph{specify in advance any arbitrary
connectivity} of the network. A particular predefined pattern is of
course not always realizable in such a network.

We illustrate this network design with predefined connectivities by
a few examples. The two small networks of Figure \ref{cap:Single-periodic-patterns}
are both networks with pre-specified absent links. Here we chose random
networks of $N=6$ neurons where each connection is present with probability
$p=0.8$. The figure displays two different networks
that exhibit the same pattern. One network has been chosen such that
the pattern is stable the other such that it is unstable. Interestingly,
on the one hand the same pattern can be invariant in two different
networks with similar statistics, on the other hand their stability
properties depend on the details of the coupling configurations.

We also considered large networks by predefining exactly the presence or
absence of each link
according to very different degree distributions. We designed them,
by varying the remaining (non-zero) coupling strengths, such that
all network examples exhibit the same predefined simple-periodic pattern.
Network design on specific connectivities is of course not restricted 
to the example cases presented here,
because the sets of input coupling strengths can be specified independently
from each other.

For illustration, we present four large networks of $N=1000$ neurons
realizing the same predefined periodic pattern of spikes. For simplicity,
we took for all networks the in-degree equal to the out-degree for
each neuron. A random degree sequence was drawn from the given degree
distribution (see below) and the degrees assigned to the neurons. 
The networks were then generated using a
Monte-Carlo method similar to those discussed in Ref.\ \cite{MKINA03}.

 Approximately $50$\% of the neurons are of integrate-and-fire
type, the remaining are of Mirollo-Strogatz-type. The parameters of
the leaky integrate-and-fire neurons are randomly chosen within
$I_{m}\in(1.08,2.08)$, $\gamma_{m}\in(0.5,1.5)$,
the parameters $b_{m}$ of the Mirollo-Strogatz neurons are randomly
chosen in $b_{m}\in(0.9,1.2)$, then $a_{m}\in(1/(e^{b_{m}}-1)-0.1,1/(e^{b_{m}}-1)+0.1)$.
The thresholds of both neuron types are uniformly distributed within
the interval $\Theta_{l}\in(0.8,1.2)$. The delay distribution is
heterogeneous, delays are uniformly distributed in the interval
$\tau_{lm}\in (0.1,0.3)$, $l,m\in\{1,...,N\}$.

\begin{figure}[b]
\begin{center}\includegraphics{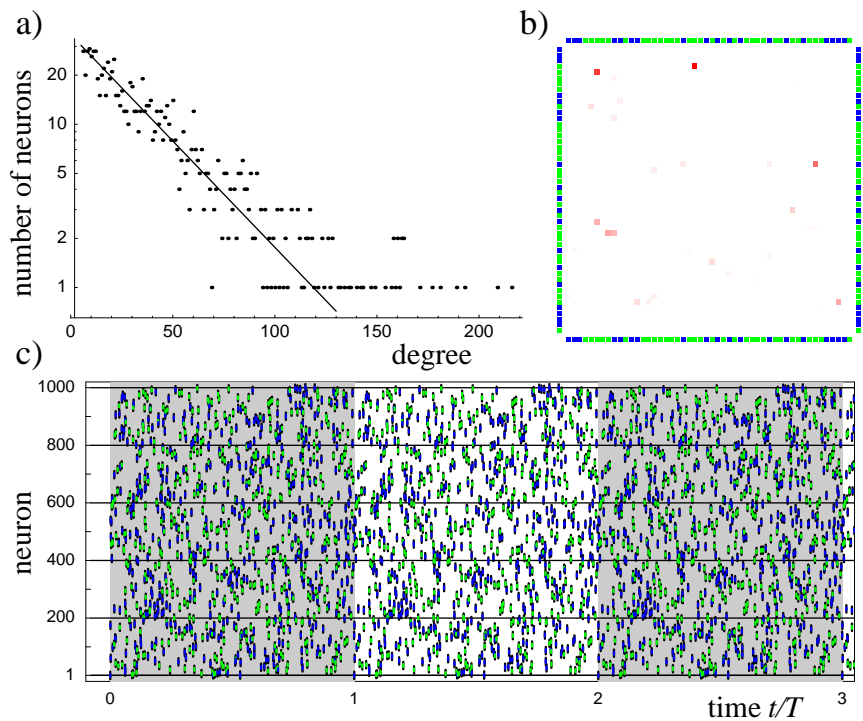}\end{center}
\caption{(color) Network design with given connectivity. Predefined pattern
in a network ($N=1000$) with exponential degree distribution (panel
(a), $\alpha=0.03$) and purely inhibitory coupling. Panel (b) displays
the sub-matrix of coupling strengths between the first $50$ neurons.
Inhibitory couplings are red, excitatory couplings are gray. The intensity
of the color is proportional to the coupling strength. Due to too faint
color, some very weak couplings are invisible in the plot. The frame shows
integrate-and-fire neurons in green and Mirollo-Strogatz neurons in
blue. (c) The numerical simulations of the designed networks (green
and blue bars for integrate-and-fire and Mirollo-Strogatz type neurons)
show perfect agreement with the predefined pattern (black bars).\label{cap:Expo0.5inh}}
\end{figure}

\begin{figure}[p]
\begin{center}\includegraphics{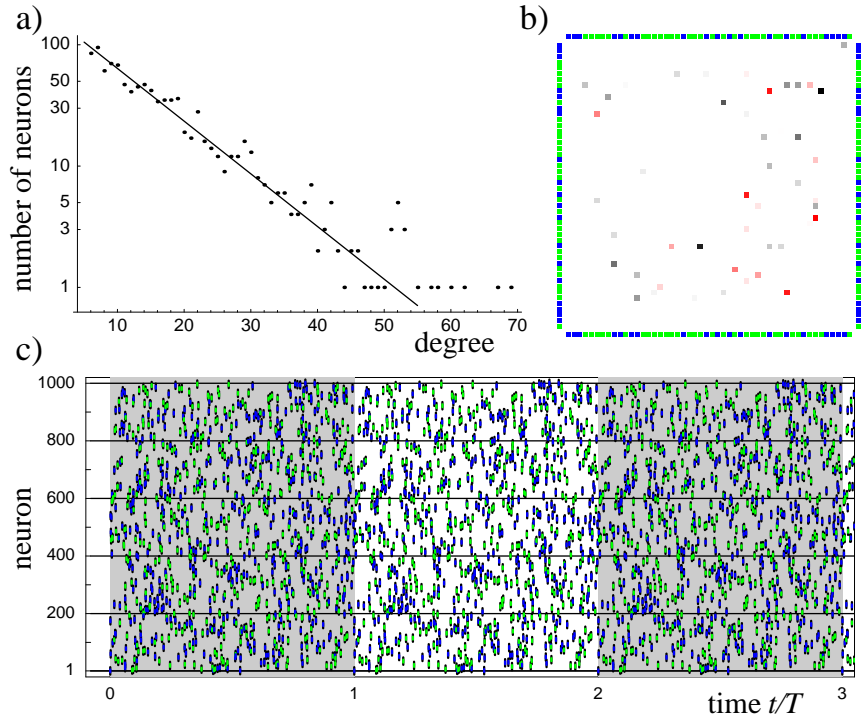}\end{center}
\caption{(color) Network design with given connectivity. Predefined pattern
in a network ($N=1000$) with exponential degree distribution (panel
(a), $\alpha=0.1$) and mixed inhibitory and excitatory coupling.
Other panels as in Figure \ref{cap:Expo0.5inh}.\label{cap:Expo0.5inhexc}}
\end{figure}

\begin{figure}[p]
\begin{center}\includegraphics{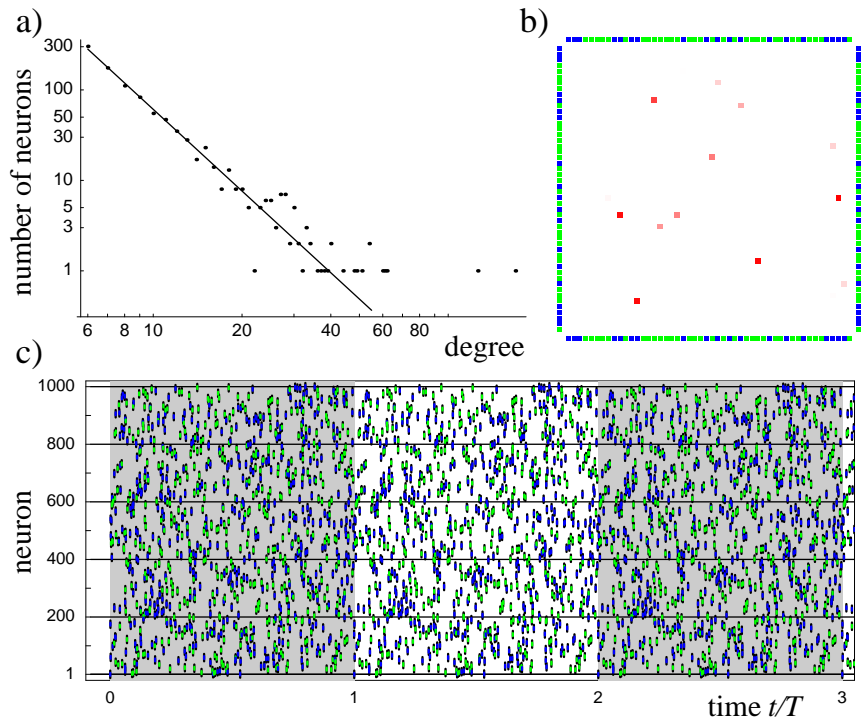}\end{center}
\caption{(color) Network design with given connectivity. Predefined pattern
in a network ($N=1000$) with power-law degree distribution (panel
(a), $\gamma=3.0$) and purely inhibitory coupling. Other panels as
in Figure \ref{cap:Expo0.5inh}.\label{cap:ScaleFree3.0inh}}
\end{figure}

\begin{figure}[t]
\begin{center}\includegraphics{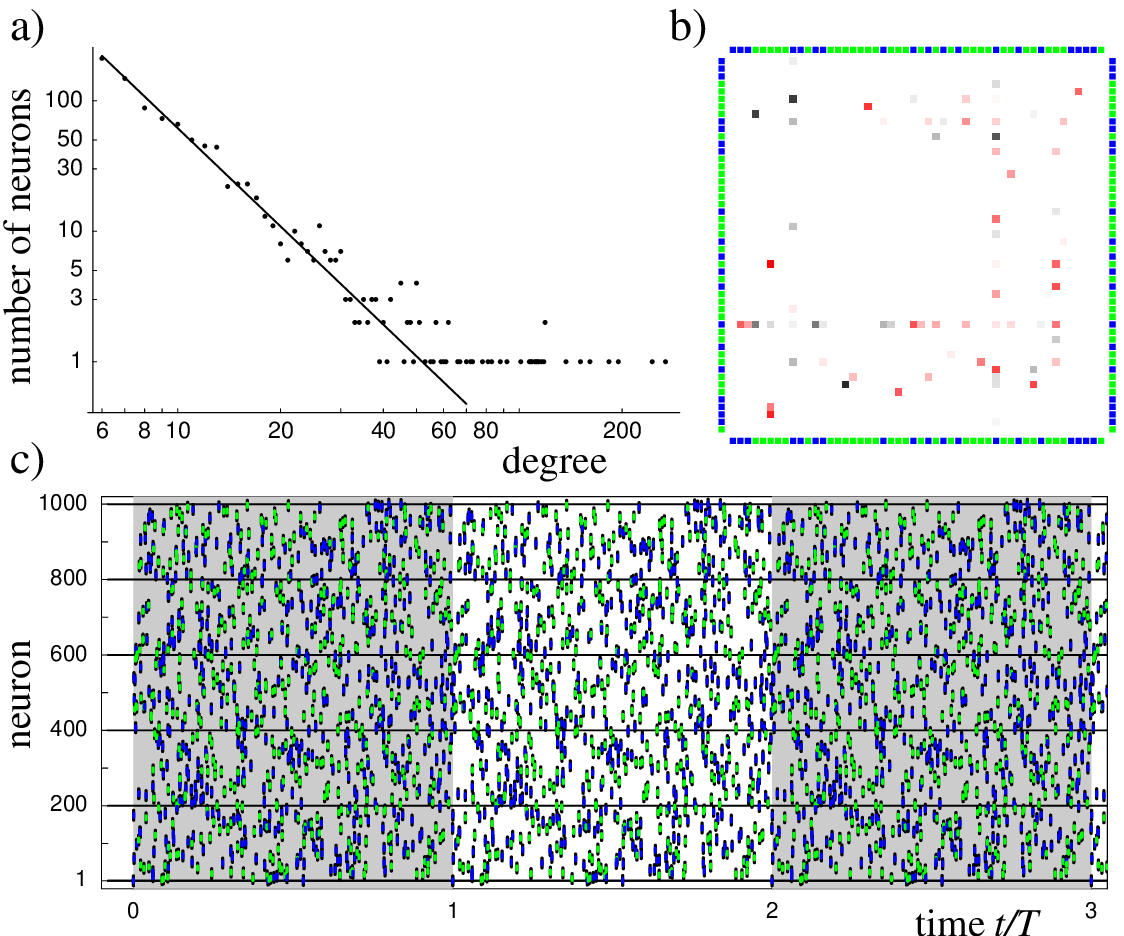}\end{center}
\caption{(color) Network design with given connectivity. Predefined pattern
in a network ($N=1000$) with power-law degree distribution (panel
(a), $\gamma=2.5$) and mixed inhibitory and excitatory coupling.
Other panels as in Figure \ref{cap:Expo0.5inh}.\label{cap:ScaleFree2.5inhexc}}
\end{figure}

Two network examples (Figs. \ref{cap:Expo0.5inh},\ref{cap:Expo0.5inhexc})
have random connectivity with different exponential degree distributions\begin{equation}
p(k)\propto e^{-\alpha k}\label{eq:EXPOdegree}\end{equation}
 where $k$ is the neuron degree. The other two networks (Figs. \ref{cap:ScaleFree3.0inh},\ref{cap:ScaleFree2.5inhexc})
have power-law degree distribution, according to \begin{equation}
p(k)\propto k^{-\gamma}\label{eq:SFdegree}\end{equation}
For both distributions, we fixed a lower bound on the degree $k_{c}=6$
such that each neuron has $k\geq k_{c}$ input and output connections.
For networks of both distributions, we realized one with purely inhibitory
coupling strengths (Figs. \ref{cap:Expo0.5inh},\ref{cap:ScaleFree3.0inh})
and one with mixed inhibitory and excitatory coupling strengths (Figs.
\ref{cap:Expo0.5inhexc},\ref{cap:ScaleFree2.5inhexc}).

All network examples are constructed to realize the same predefined
spike pattern with period $T=1.5$. The numerical simulations (Figs.
\ref{cap:Expo0.5inh}-\ref{cap:ScaleFree2.5inhexc}c, green or blue
bars for spiking integrate-and-fire or Mirollo-Strogatz-type neurons)
agree perfectly with the predefined pattern (Figs. \ref{cap:Expo0.5inh}-\ref{cap:ScaleFree2.5inhexc}c,
underlying black bars).
\begin{rem}
Due to the simplicity of imposing absence of links, the same method
can be applied to a wide variety of network connectivities. In particular,
a connectivity can be randomly drawn from any kind of degree distribution;
a connectivity can also be structured (e.g. correlated degrees) and
one may want to implement a very detailed specific form of it, e.g.,
as given by real data.
\end{rem}
As noted above, however, not all networks can be designed for any
pattern; in particular it is in general necessary to have sufficiently
many incoming links to each neuron such that the interaction delay
times and the input coupling strengths can account for the desired
phase dynamics consistent with the predefined spike pattern.

\section{Designing optimal networks}
\label{sec:OptimalNetworks}

In section \ref{sec:NetworkDesign} we derived analytical constraints
specifying the set of all networks that exhibit a predefined pattern
and found that often there is a multi-dimensional family of solutions
in the space of networks (as defined by all coupling strengths). In
the previous section we exploited this freedom to design networks
the connectivity of which is specified in detail. We may also exploit
the freedom of choosing a solution among many possibilities by optimizing
certain network properties.

Can we design networks that optimize certain structural features 
and at the same time exhibit a predefined pattern dynamics? This question
is a very general one and it can be addressed by considering a variety
of features of neuroscientific or mathematical interest. To
briefly illustrate the idea, we here focus on optimizing convex 
'cost' functions of the coupling strengths $\eps_{lm}$ and 
look for those networks among the admissible ones that minimize wiring costs.

Even for this very specific problem there are a number of different
approaches we can take. For instance, we can consider networks
with the same type of interactions, inhibitory or excitatory, or allow
for a mixture of both, or optimize for different features of the connectivity. 
For simplicity, we here consider small networks whose neurons are exclusively of integrate-and-fire
type and allow for a mixture of inhibitory and excitatory coupling.
Integrate-and-fire neurons have the advantage (for both analysis and
optimization) that the constraints (\ref{eq:F2A})--(\ref{eq:F2D})
are linear.

\begin{figure}[tb]
\begin{center}\includegraphics[%
  width=0.8\textwidth]{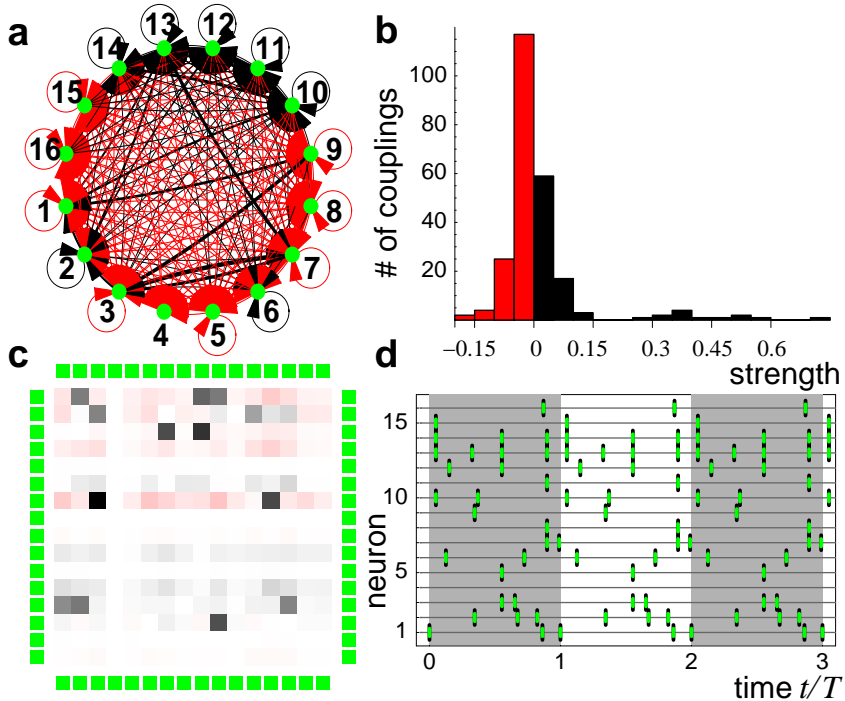}\end{center}
\caption{(color) Network of leaky integrate-and-fire neurons that minimizes
the wiring cost in Euclidean norm by minimizing (\ref{eq:2NormOpt}). The parameters
are randomly chosen within $I_{m}\in(1.0,2.0)$, $\gamma_{m}\in(0.5,1.5)$
and $\Theta_{m}\in(0.8,1.2)$. The delays are uniformly distributed
in $\tau_{lm} \in (0.1,0.9)$, $l,m\in\{1,...,N=16\}$. Panels (a) and (c)
show the network and the coupling matrix $\eps_{lm}$. Panel (b) shows
the histogram of the strengths of existing connections in the network.
The bin size is 0.05. Panel (d) displays the predefined spike pattern
(black bars) that is accurately reproduced (green bars). In the optimal
network every neuron is connected to every other except the silenced
neuron $l=4$. This neuron has no outgoing connections: Since it generates
no spikes, outgoing connections would be superfluous and do not appear in the
optimal network. \label{cap:OptEuclNW}}
\end{figure}
The most straightforward goal for optimizing wiring costs is to minimize
the quadratic cost function
\begin{equation}
G(\eps):=\sum_{l=1}^{N}\sum_{m=1}^{N}\eps_{lm}^{2}\,,\label{eq:2NormOpt}
\end{equation}
A similar approach has already been successfully used when minimizing
wiring costs of biological neural networks based on anatomical and
physical constraints but neglecting dynamics issues, see, e.g. \cite{C04}.
When minimizing the Euclidian ($L_2$) norm $\sqrt{G(\eps)}$ by minimizing (\ref{eq:2NormOpt}) for each 
row vector $(\eps_{l,m})_{m\in \{1,\ldots,N\}}$ of the coupling matrix, a solution
is searched among the admissible ones that is closest to the origin
in the space of networks (defined by the coupling strengths).

Figure \ref{cap:OptEuclNW} shows an example of such an optimization.
The network is almost globally connected and shows moderate variation
among the individual coupling strengths. The predefined pattern dynamics
is exactly reproduced. Such a network, while optimizing the wiring
cost according to (\ref{eq:2NormOpt}) does not appear to have any
special features apart from apparently homogeneous and relatively
small coupling strengths. %

\begin{figure}[tb]
\begin{center}\includegraphics[%
  width=0.8\textwidth]{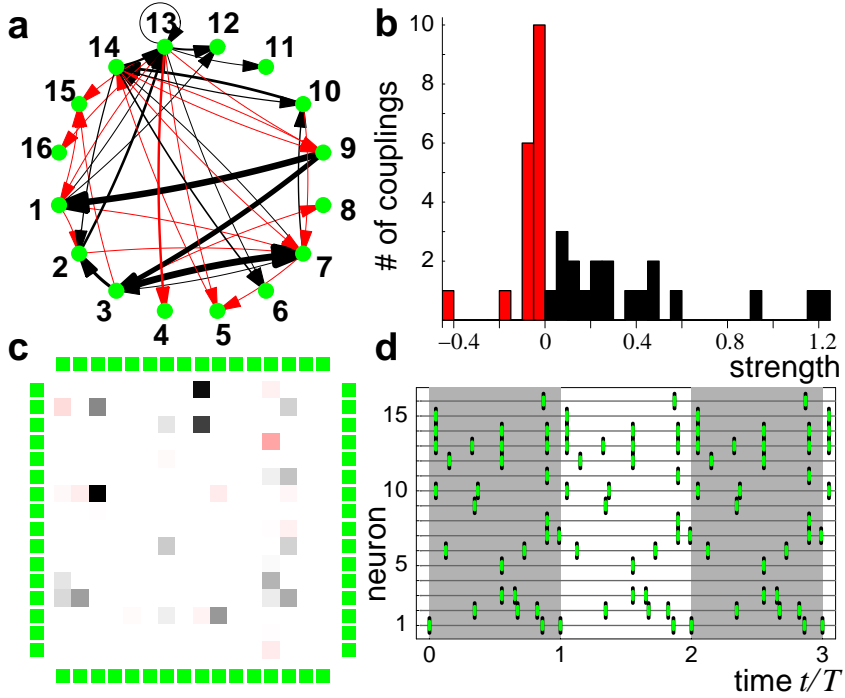}\end{center}
\caption{(color) Network that minimizes the wiring cost in $L_{1}$-norm (\ref{eq:1NormOpt}).
The parameters are randomly chosen within $I_{m}\in(1.0,2.0)$, $\gamma_{m}\in(0.5,1.5)$
and $\Theta_{m}\in(0.8,1.2)$. The delays are uniformly distributed
in $\tau_{lm} \in (0.1,0.9)$, $l,m\in\{1,...,N=16\}$. Panels (a) and (c)
show the network and the coupling matrix $\eps_{lm}$. Panel (b) shows
the histogram of the strengths of existing connections in the network.
The bin size is 0.05. Panel (d) displays the predefined spike patterns
(black bars) that is accurately reproduced (green bars). The optimal
network is very sparsely connected. In fact the network has one large
strongly connected component, containing the neurons $\{1,2,3,5,7,9,10,13,14\}$,
while the remaining neurons receive connections exclusively from this
component and do not have any outgoing connections. \label{cap:OptAbsNW}}
\end{figure}
It seems 
that nature often designs networks in a different way,
possibly such that they serve a dynamical purpose especially well. 
In particular evolution has not
optimized most biological neural networks in the above manner: they are not
close to globally coupled.

An alternative goal for optimizing wiring costs is to minimize the
cost function\begin{equation}
G(\eps):=\sum_{l=1}^{N}\sum_{m=1}^{N}\left|\eps_{lm}\right|,\label{eq:1NormOpt}\end{equation}
 that is, the $L_{1}$-norm of each row vector of the coupling matrix. 
When minimizing the $L_{1}$-norm (\ref{eq:1NormOpt}), as before, 
a solution is searched among the admissible ones that is closest to the origin in the space of networks, but this time 'close' is defined by the $L_{1}$ distance
measure. Interestingly, under weak conditions on the linear equality
constraints, an optimal solution (\ref{eq:1NormOpt}), searched under
these constraints only, has many entries $\eps_{lm}$ equal to zero,
cf. \cite{BV04}. Because we typically also have many inequalities
which depend on details of the pattern dynamics and are therefore
uncontrolled, we cannot guarantee the zero entries for the full optimization
problem (defined by equalities and inequalities) here. However, our
numerics suggests that the solution in fact gives a network with many
links absent and the number of links present being typically of the order
of number of \emph{equality} constraints.

Thus a network optimized by minimizing the $L_{1}$-norm is sparse,
see, e.g., Fig. \ref{cap:OptAbsNW}. Moreover, compared to the optimal
$L_{2}$-norm solution above, this network has more heterogeneous
connection strengths. Given some type of dynamics, a sparse network
possibly is what biological systems would optimize for. In biological
neural networks for instance, creating an additional synapse would
probably use more resources (energy, biological matter, space, time,
etc.) than making an existing synapse stronger.

Sparseness might possibly also be optimized in biological neural networks
where requirements are met enabling other specific, functionally relevant
dynamics. In general, of course, this dynamics may or may not consist
of spike patterns.

\begin{rem}
The optimization problem, (\ref{eq:2NormOpt}) and (\ref{eq:1NormOpt})
with constraints (\ref{eq:F2A})--(\ref{eq:F2D}), \textbf{}does typically
not have a true optimum.
\end{rem}

If a pattern is predefined that has more than one reception times
between two successive sending events of some neuron, there
usually are strict inequalities among the constraints (\ref{eq:F2A})--(\ref{eq:F2D}). 
Because the functions $H_{\eps}^{(l)}$ in (\ref{eq:F2A})--(\ref{eq:F2D})
are local homeomorphisms (i.e. are continuous with local inverses
that are continuous) the set of admissible coupling strengths is then not
closed and thus does not contain its boundary.

During optimization, typically a solution is searched that is as close
to such a boundary as possible.
For instance, suppose one connection from $m$
to $l$ is inhibitory and its strength $\eps_{lm}$ is desired as
small as possible. Then a solution is searched where the phase $\phi_{l}$
of the neuron $l$ that receives a spike from $m$ is such that the
phase jump that spike induces is maximal (in absolute value) when $\eps_{lm}$
is held constant. This
way a given desired phase jump would be achieved by a minimal coupling
strength. Typically, the phase $\phi_{l}$ sought-after corresponds to
a boundary of the set of admissible phases. For instance, if $U_l$ is concave,
an inhibitory spike has the largest possible effect  on $\phi_{l}$
(largest phase jump) at $\phi_{l}=\Theta_{l}$. The corresponding phase
constraint, however, may read $\phi_{l}<\Theta_{l}$. Thus the boundary
phase and therefore also the boundary coupling strength cannot be
assumed. As a consequence, the optimization problem has no true solution.

We fix this problem by imposing, instead of (\ref{eq:F2A})--(\ref{eq:F2D})
\textbf{}and possible additional constraints with \emph{inequalities}
of the type $\phi_{l}>x$ or $\phi_{l}<y$, constraint sets that are
closed, i.e. $\phi_{l}\geq x+\kappa$ or $\phi_{l}\leq y-\kappa$,
where $\kappa>0$, $\kappa\ll1$ is a small cutoff. We fixed $\kappa=0.001$
\textbf{}in the optimal design problems considered here.

\section{Brief Network Design Manual}
\begin{figure}[htb]
\begin{center}\includegraphics[%
  width=75mm]{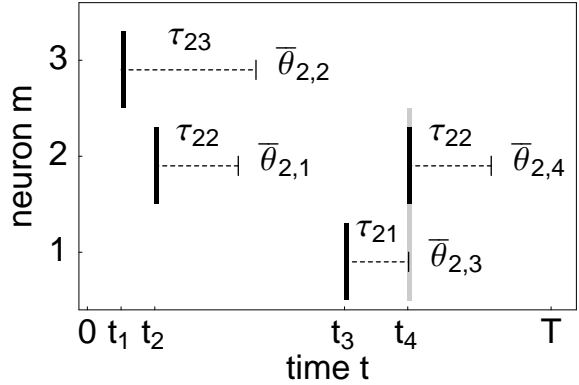}
\end{center}

\caption{Pattern of $M=4$ spikes in a network of $N=3$ neurons illustrating
the indexing of spike sending and reception times. The spike (sending) times
$t_i$, marked by black bars, are indexed with increasing $i$ according to
their temporal order of occurrence in the network (the neuron identities play no role for this
index). The ordered spike reception times $\bar{\theta}_{l,i}$ are displayed
for neuron $l=2$. They are generally different for other receiving neurons
($l\neq 2$, not shown) and obtained by adding the delay times $\tau_{lm}$ (dashed lines) to the spike sending times $t_j$ and then ordering the resulting set for each neuron. Here there is one degenerate event: neuron $l=2$ receives a spike from $m=1$ exactly at its second spike sending time $t_4$ (light gray vertical bar).
\label{cap:threeneurons}
}
\end{figure}

In this section we briefly summarize the presented method (of designing the coupling strengths of a network such that it realizes a pre-defined pattern) by providing step-by-step instructions. For simplicity, as above, we assume that all other parameters, such as neuron rise functions and interaction delay times are given or fixed a priori. We refer to the relevant sections and formulas derived above where appropriate. A simple example of a small network of $N=3$ neurons (Fig.~\ref{cap:threeneurons}) illustrates the indexing used in the general instructions.

Suppose a periodic pattern of $M$ spikes is given in a network of $N$ neurons.

1) Label the neurons arbitrarily by $m\in \{1,\ldots,N\}$.

2) Fix the origin of time, $t=0$, arbitrarily and pick an interval of length $T$, the period of the given pattern. 

3) Order the spike times. Some neurons may send one spike per period, others
   multiple spikes, and again others no spike at all (silent neuron). Label
   the times of all spike sending events according to their temporal order of occurrence in the network. In the example  of Fig.~\ref{cap:threeneurons}, we have one
   spike time $t_1$ of neuron $m=3$, two spike times $t_2$ and $t_4$ of neuron
   $m=2$ and one spike time $t_3$ of neuron $m=1$.

4) Compute the spike reception times at each neuron $l$ using the interaction delay times $\tau_{lm}$ such that $\theta_{l,j}=t_j+\tau_{lm}$. Here $m$ is that neuron that sent the spike at time $t_j$.  
We identify this neuron by $s_j:=m$ in the formulas above. For those neurons $l$ for which the spike reception times are not ordered, reorder them by permuting indices according to (\ref{eq:OrderedArrivalTimes}) to obtain ordered reception times $\bar{\theta}_{l,j}$. 
In the example, the delay time $\tau_{23}$ from neuron $m=3$ to neuron $l=2$,
is longer than $\tau_{22}$, which, for the given
pattern, results in reception times $\theta_{2,j}$ that are not in the same
order as the spike sending times $t_j$. Particularly we have 
$\bar{\theta}_{2,1}=\theta_{2,2}$, $\bar{\theta}_{2,2}=\theta_{2,1}$, $\bar{\theta}_{2,3}=\theta_{2,3}$ and  $\bar{\theta}_{2,4}=\theta_{2,4}$. 
The ordered reception times $\bar{\theta}_{2,j}$ are as indicated in Figure \ref{cap:threeneurons}.

5) Are there degenerate times at which a reception time at one neuron equals
 that neuron's spike sending time? If so, decide whether to use, for each such
 reception, supra-threshold or sub-threshold input signals; for each non-degenerate spike reception, use sub-threshold inputs. In the example, the time at which neuron $2$ receives a spike from neuron $1$ coincides with the second spike sending time $t_4=\bar{\theta}_{2,3}$ of neuron $2$. So for this reception time $\bar{\theta}_{2,3}$ of neuron $l=2$, decide whether to use sub- or supra-threshold input. For all other receptions at neuron $l=2$, use sub-threshold input.

6) For each neuron $l$ and each spike time $t_k$ of that neuron, look for the previous spike time of neuron $l$ and name it ``$t_i$''. Compute and look up the particular response functions $H^{(l)}_\eps$, the thresholds $\Theta_l$ and the differences in spike reception times $\Delta_{l,j}$. 
Now, if there is 

\begin{itemize}
\item[(a)] no spike reception at time $t_i$ and no supra-threshold input generating $t_k$ write down system (\ref{eq:F2A}). 
\item[(b)] a spike reception at $t_i$ inducing the spike at $t_i$ by a
  supra-threshold input and no supra-threshold input generating $t_k$, write down system (\ref{eq:F2B}).
\item[(c)]  a spike reception at time $t_i$ but the threshold is nevertheless
  reached by the neuron from its intrinsic dynamics (as desired by the
  designer) and no supra-threshold input generating $t_k$ : if the coupling,
  effective after reset at $t_i$, is (i) subthreshold, this is a special case of (\ref{eq:F2A}); (ii) if it is supra-threshold, supplement (\ref{eq:F2B}) with (\ref{eq:F2BAddCond}).
\item[(d)] case (a) with supra-threshold input generating $t_k$ write down  (\ref{eq:F2A}) with the equation replaced by (\ref{eq:F2C}).
\item[(e)] case (b) with supra-threshold input generating $t_k$ write down
  (\ref{eq:F2B}) and replace the equation by (\ref{eq:F2D}).
\item[(f)] (i) for the case (c,i) with supra-threshold input generating $t_k$, write down
  (\ref{eq:F2A}) and replace the equation by (\ref{eq:F2C}) (ii) for the case (c,ii)
  write down (\ref{eq:F2B}) completed by (\ref{eq:F2BAddCond}) and replace the
  equation by (\ref{eq:F2D}).
\end{itemize}
Repeat this step 6) for all neurons $l$ and all pairs $(t_i,t_k)$ of their successive spike times.
  
At this point, a complete list of restricting equations and inequalities has been created. One particular solution to these restrictions provides all coupling strengths of a network that exhibits the predefined pattern as an invariant dynamics. The set of all solutions thus provides the set of all networks that exhibit this spike pattern.

One can now either 

7) solve for one particular solution; or

8) further restrict the constraint system, e.g. by requiring additional
   properties of the connectivity, cf. section \ref{sec:AdditionalRequirements}, and solve that for a particular solution; or

9) use the entire constraint system and try to find a solution that is optimal in a desired sense, as done in section 
\ref{sec:OptimalNetworks} for the example of minimal wiring costs; or

10) combine additional restrictions, point 8), and optimization, point 9).

Point 10) has not been presented in this manuscript but is an interesting starting point for future research. 
 
We found it useful to start trying these network design methods on small network examples of simple units, for instance integrate-and-fire neurons, and investigate very simple patterns with few (or no) degeneracies first. Moreover, given that there is no general recipe about how to apply additonal restrictions and how to solve general optimization problems, it might also be useful to start with few restrictions and simple optimization tasks in \emph{very} small networks the dynamics of which (and possibly their desired ``optimal'' features) can be understood intuitively.

\section{Conclusions}

\subsection{Summary}

In this article, we have shown how to design model networks of spiking
neurons such that they exhibit a predefined dynamics. We focused on
the question of how to adapt the coupling strengths in the network
to fix the dynamics. We derived analytical constraints on the coupling
strengths (which define the set of all networks) given an arbitrarily
chosen predefined periodic spike pattern. The analysis presented here
is very general. It covers networks of arbitrary size and of different
types of neurons, heterogeneously distributed delays and thresholds
(and thus intrinsic neuron frequencies), combinations of inhibitory
and sub- and supra-threshold excitatory interactions as well as complicated
stored patterns that include degenerate event times, multiple spiking
of the same neuron within the pattern and silent neurons that never
fire. These constraints do not admit a solution for certain
patterns. Once the features of individual neurons and the delay-distribution
are fixed, this implies that these patterns cannot exist in any network,
no matter how the neurons are interconnected.

A predefined simple periodic pattern is particularly interesting because
under weak assumptions, the constraint system has a solution for any
such pattern. Thus, a network realizing any simple periodic pattern
is typically guaranteed to exist; we analytically parameterized all
such networks. The family of solutions is typically high-dimensional, cf. also
\cite{PBM04},
and we showed how to design networks that are further constraint.
We highlighted the possibility to design networks of completely predetermined
connectivity (fixing the absence or presence of links between each
pair of neurons). To illustrate the idea, we have explicitely designed
networks with different exponential and power-law degree distributions
such that they exhibit the same spike pattern. 

The design perspective can furthermore be used to find networks that
exhibit a predefined dynamics and are at the same time optimized in
some way. As a first example, we considered networks minimizing
wiring cost. The connectivity of biological neural networks that exhibit 
precise spatio-temporal spiking dynamics is typically sparse. The work 
presented here suggests that this sparseness may result from an optimization 
process that takes into account dynamical aspects. If biological neural 
networks indeed optimize connectivity for dynamical purposes, our results 
suggest that these networks may minimize the total number of 
connections (rather than, e.g., their total strengths) and at the same time
still realize specific spiking dynamics.

\subsection{Perspectives for future research}

The dynamics of artifically grown biological neural networks may provide
an immediate application ground for the theory presented here. For
instance, to uncover the origin of recurring, specific spike patterns,
one could imagine using a design approach to precisely control the
growth of biological neural networks on artificial substrates and
reveal under which conditions and how a desired pattern arises in
a biological environment. For practicability of such an approach,
of course, pattern stability, only briefly discussed here, needs a
more detailed analysis. Moreover, the size of the basin of attraction
of a spike pattern will probably also play an important role in such
studies. Perhaps it may even become possible to develop design techniques
to optimize pattern stability and basin size, thus gaining robust
pattern dynamics.

Network design might be a valuable new perspective of research, as
shown here by example for spiking neural networks. Using the design
idea might not only aid a better understanding of the relations between
structure and function of complex networks in general; network design
might also be exploited for systems that we would like to fulfill
a certain task, for example computational systems such as artificial
neural networks. 

The idea of designing a system of coupled units is not new. For instance
an artificial Hopfield neural network \cite{H82} can be trained by
gradually adapting the coupling strengths, such that it becomes an
associative memory, fulfilling a certain pattern recognition task.
Such networks typically consist of binary units that are all-to-all
coupled. However, already in the late 1980's \cite{DGZ87} mean field
theory has been successfully extended to study the properties of sparse,
randomly diluted Hopfield networks. In that work, Derrida, Gardner
and Zippelius showed that the storage capacity of such diluted systems
is reduced compared to the all-to-all coupled one, but still significant.

Here we transferred the idea of system design to complex networks
that may have a complicated, irregular connectivity and thus cannot
in general be described by mean field theory. 
In related study \cite{MPF05}, a method has been presented to construct neural
network
models that exhibit spike trains with high statistical correlation to given
extracellular recordings.
The specific results
presented our this study might be valuable to obtain further insights
into biological neural systems and the precisely timed, still unexplained,
spike patterns they exhibit. This study, however, also raises a number
of questions both for the theory of spiking neural network as well
as, more generally, for studies of other complex networks and their
dynamics. We list a few questions we believe are among the most interesting,
and promising in the near future:

Can network design studies help to develop functionally relevant dynamics?
Design of particular model networks could on the one hand identify
possible functional (as well as irrelevant) subgroups of real-world
networks, including neural, gene and social interaction networks;
on the other hand network design could also guide the development
of new useful paradigms and devices, for instance for information
processing or communication networks.

What is an optimal network design that ensures synchronization \textbf{}\cite{PRK01},
a prominent kind of collective dynamics? The approach could of course
also be useful to avoid certain behavior. For instance, may network
design even give hints about how to suppress synchronization and hinder
epileptic seizures in the brain (see e.g.~\cite{PHT06} and references
therein)? What are potential ways to design your favorite network?
What kind of dynamics would be desirable (or undesirable$^{*}$) for
it. 

Let's use network design -- and make specific network dynamics (not$^{*}$)
happen.

\begin{ack}
We thank Sven\ Jahnke, Michael\ Herrmann, Fred\ Wolf and Theo\ Geisel
for inspiring discussions; MT thanks the researchers, staff and students
at the Center for Applied Mathematics, and at the Department for Theoretical
and Applied Mechanics, Cornell University, in particular, Steven Strogatz
and Richard Rand, for kind hospitality and an enjoyable work atmosphere;
MT further thanks Levke Johanna Deutsch and Benika Pinch for carefully
reading a report related to this article. 

Both authors acknowledge partial support from the Ministry for Education
and Science, Germany, via the Bernstein Center for Computational Neuroscience,
Göttingen, under grant number 01GQ0430. MT acknowledges financial
support by the Max Planck Institute for Dynamics and Self-Organization
and the Max Planck Society through a scholarship based on his award
of the Otto Hahn Medal. 
\end{ack}
\bibliographystyle{abbrv}

\begin{thebibliography}{10}

\bibitem{A82}
M.~Abeles.
\newblock {\em Local Cortical Circuits: An Electrophysiological Study}.
\newblock Springer, Berlin, 1982.

\bibitem{A04}
M.~Abeles.
\newblock Time is precious.
\newblock {\em Science} 304:523, 2004.

\bibitem{AY92}
T.~Achacoso and W.~Yamamoto, editors.
\newblock {\em AY's Neuroanatomy of C. elegans for computation}.
\newblock CRC Press, 1992.

\bibitem{AT05}
P.~Ashwin and M.~Timme.
\newblock Unstable attractors: existence and robustness in networks of
  oscillators with delayed pulse coupling.
\newblock {\em Nonlinearity} 18:2035,  2005.

\bibitem{AMAH03}
Y.~Aviel, C.~Mehring, M.~Abeles, and D.~Horn.
\newblock On embedding synfire chains in a balanced network.
\newblock {\em Neural Comput.} 15:1321, 2003.

\bibitem{DGZ87}
A.~Zippelius,  B.~Derrida, E.~Gardner.
\newblock An exactly solvable asymmetric neural network model.
\newblock {\em Europhys. Lett.} 4:167, 1987.

\bibitem{BV04}
S.~Boyd and L.~Vandenberghe.
\newblock {\em Convex Optimization}.
\newblock Cambridge Univ. Press, Cabridge, UK, 2004.

\bibitem{C04}
D.~B. Chklovskii.
\newblock Exact solution for the optimal neuronal layout problem.
\newblock {\em Neural Comput.} 16:2067, 2004.

\bibitem{DTDWG04}
M.~Denker, M.~Timme, M.~Diesmann, F.~Wolf, and T.~Geisel.
\newblock Breaking synchrony by heterogeneity in complex networks.
\newblock {\em Phys. Rev. Lett.} 92:074103, 2004.

\bibitem{DGA99}
M.~Diesmann, M.-O. Gewaltig, and A.~Aertsen.
\newblock Stable propagation of synchronous spiking in cortical neural
  networks.
\newblock {\em Nature} 402:529, 1999.

\bibitem{DK02}
B.~Drossel and A.~McKane.
\newblock {\em Modelling Food Webs}.
\newblock Wiley-VCH, 2002.

\bibitem{EPG95}
U.~Ernst, K.~Pawelzik, and T.~Geisel.
\newblock Synchronization induced by temporal delays in pulse-coupled
  oscillators.
\newblock {\em Phys. Rev. Lett} 74:1570, 1995.

\bibitem{EPECM99}
C.~W. Eurich, K.~Pawelzik, U.~Ernst, J.~D. Cowan, and J.~G. Milton.
\newblock Dynamics of self-organized delay adaptation.
\newblock {\em Phys. Rev. Lett.} 82:001594, 1999.

\bibitem{GS05} 
K.~Gansel and W.~Singer.
\newblock Replay of second-order spike patterns with millisecond precision in
  the visual cortex.
\newblock {\em Soc. Neurosci. Abstr.} 276.8 (2005).

\bibitem{HHP95}
M.~Herrmann, J.~A. Hertz, and A.~Pr\"ugel-Bennett.
\newblock Analysis of synfire chains.
\newblock {\em Network} 6:403, 1995.

\bibitem{H82}
J.~J. Hopfield.
\newblock Neural networks and physical systems with emergent collective
  computational abilities.
\newblock {\em PNAS} 79:2554, 1982.

\bibitem{HMIC01}
F.~I. Jeff~Hasty, David~McMillen and J.~J. Collins.
\newblock Computational studies of gene regulatory networks: in numero
  molecular biology.
\newblock {\em Nature Rev. Genet.} 2(268), 2001.

\bibitem{J02}
D.~Z. Jin.
\newblock Fast convergence of spike sequences to periodic patterns in recurrent
  networks.
\newblock {\em Phys. Rev. Lett.} 89:208102, 2002.

\bibitem{A93}
{M. Abeles et al.}
\newblock Spatiotemporal firing patterns in the frontal cortex of behaving
  monkeys.
\newblock {\em J. Neurophysiol.} 70:1629, 1993.

\bibitem{BR02}
I.~J. {Matus Bloch} and C.~{Romero Z.}
\newblock Firing sequence storage using inhibitory synapses in networks of
  pulsatile nonhomogeneous integrate-and-fire neural oscillators.
\newblock {\em Phys. Rev. E} 66:036127, 2002.

\bibitem{MT06}
R.-M. Memmesheimer and M.~Timme.
\newblock in preparation.

\bibitem{MT06CNS}
R.-M. Memmesheimer and M.~Timme.
\newblock Spike patterns in heterogeneous neural networks.
\newblock Comp. Neurosci. Abstr. (CNS) S98, 2006.

\bibitem{MT05}
R.-M. Memmesheimer and M.~Timme.
\newblock Designing the Dynamics of Spiking Neural Networks.
\newblock Phys. Rev. Lett., 97:188101, 2006.

\bibitem{MKINA03}
R. Milo, N. Kashtan, S. Itzkovitz, M.E.J. Newman, and U. Alon.
\newblock On the uniform generation of random graphs with prescribed degree sequences.
\newblock {\em http://arxiv.org/abs/cond-mat/0312028}, 2003.


\bibitem{MS90}
R.~E. Mirollo and S.~H. Strogatz.
\newblock Synchronization of pulse coupled biological oscillators.
\newblock {\em SIAM J. Appl. Math.} 50:1645, 1990.

\bibitem{N03}
M.~E.~J. Newman.
\newblock The structure and function of complex networks.
\newblock {\em SIAM Review} 45:167, 2003.

\bibitem{PHT06}
P.~A.~Tass, O.~V.~Popovych, Christian~Hauptmann.
\newblock Control of neuronal synchrony by nonlinear delayed feedback.
\newblock Biol. Cybern., 95:69, 2006.

\bibitem{PRK01}
A.~Pikovsky, M.~Rosenblum, and J.~Kurths.
\newblock {\em Synchronization: A Universal Concept in Nonlinear Sciences}.
\newblock Cambridge Univ. Press, Cambridge, MA, 2001.

\bibitem{S99}
W.~Singer.
\newblock Neural synchrony: A versatile code for the definition of relations.
\newblock {\em Neuron} 24:49, 1999.

\bibitem{S04}
I.~Stewart.
\newblock Networking opportunity.
\newblock {\em Nature} 427:601, 2004.

\bibitem{S01}
S.~Strogatz.
\newblock Exploring complex networks.
\newblock {\em Nature} 410:268, 2001.

\bibitem{T02}
M.~Timme.
\newblock \emph{Collective dynamics in networks of pulse-coupled oscillators}.
\newblock Doctoral Thesis, Georg August University G{\"o}ttingen (2002).

\bibitem{TWG02a}
M.~Timme, F.~Wolf, and T.~Geisel.
\newblock Coexistence of regular and irregular dynamics in complex networks of
  pulse-coupled oscillators.
\newblock {\em Phys. Rev. Lett.} 89:258701, 2002.

\bibitem{TWG02b}
M.~Timme, F.~Wolf, and T.~Geisel.
\newblock Prevalence of unstable attractors in networks of pulse-coupled
  oscillators.
\newblock {\em Phys. Rev. Lett.} 89:154105, 2002.

\bibitem{TWG03}
M.~Timme, F.~Wolf, and T.~Geisel.
\newblock Unstable attractors induce perpetual synchronization and
  desynchronization.
\newblock {\em Chaos} 13:377, 2003.

\bibitem{TWG04}
M.~Timme, F.~Wolf, and T.~Geisel.
\newblock Topological speed limits to network synchronization.
\newblock {\em Phys. Rev. Lett.} 92:074101, 2004.

\bibitem{I04}
{Y. Ikegaja et al.}
\newblock Synfire chains and cortical songs: Temporal modules of cortical
  activity.
\newblock {\em Science} 304:559, 2004.

\bibitem{PBM04}
{A.~A.~Prinz, D.~Bucher, E.~Marder}
\newblock Similar network activity from disparate circuit parameters.
\newblock {\em Nature neurosci.} 7:1345, 2004.

\bibitem{MPF05}
{V.~A.~Makarov, F.~Panetsos, O.~de Feo}
\newblock A method for determining neural connectivity and inferring the
underlying network dynamics using extracellular spike recordings.
\newblock {\em J. Neurosci.Meth.} 144:265, 2004.

\end{thebibliography}

\end{document}